\documentclass[11pt]{article}
\usepackage{graphicx}
\usepackage{color}
\usepackage{amsmath,amssymb,amsfonts,latexsym}
\usepackage{makecell}
\begin{document}

\begin{center}
\textbf{\Large Gaussian Process Modeling of Time Series}

\vspace{7mm}
{\large Genshiro Kitagawa\\[3mm]

Tokyo University of Marine Science and Technology\\
and\\
The Institute of Statistical Mathematics
}

\vspace{3mm}
{\today}
\end{center}

\vspace{2mm}

\begin{center}{\bf\large Abstract}\end{center}

\begin{quote}
Gaussian processes (GPs) provide a flexible nonparametric framework for modeling time series through appropriately chosen kernel functions. This chapter introduces the basic formulation of Gaussian processes, commonly used kernels, GP regression, hyperparameter estimation, and model evaluation using in-sample and out-of-sample criteria. Applications to stationary, quasi-periodic, and seasonal time series illustrate how individual and composite kernels can represent different forms of temporal variation. Additive kernels also provide interpretable decompositions into latent components such as trend, smooth local variation, and seasonality, while product kernels allow more complex dependence structures to be constructed. Finally, Gaussian process state-space models (GP-SSMs) are briefly introduced, and a nonlinear example demonstrates how a GP transition model can be combined with particle filtering and smoothing for latent-state estimation.

\end{quote}

\vspace{5mm}

\section{Introduction}

Gaussian processes (GPs) provide a flexible framework for modeling unknown functions and have been widely used for regression, prediction and nonlinear modeling. A major advantage of Gaussian process regression is that complex functional relationships can be represented through covariance, or kernel, functions without specifying a particular parametric form. The choice of kernel determines important properties of the resulting process, such as smoothness, periodicity, and long-range dependence, and combinations of kernels make it possible to represent several types of variation simultaneously.

This report provides an introductory account of Gaussian process modeling with particular emphasis on time series applications\cite{RW 2006,Roberts 2013}. We first review the basic formulation of Gaussian processes and several commonly used kernel functions, including the rbf, exponential, linear, periodic, rational quadratic and Matérn kernels. Gaussian process regression and the estimation of kernel hyperparameters by maximum likelihood are then described. In addition to the in-sample marginal likelihood and AIC, the out-of-sample predictive likelihood is considered for evaluating the predictive performance of fitted GP models.

The use of different kernels and their combinations is illustrated through three time series examples having different characteristics. A stationary ship roll motion series is used to examine the ability of GP regression to reconstruct a time series from a relatively small number of observations. Annual sunspot data illustrate the use of kernels for quasi-periodic variation, including additive and product kernels. A seasonal economic time series is then used to demonstrate how an additive combination of linear, rbf and periodic kernels can represent trend, smooth local variation and seasonality as separate latent components.

Finally, we briefly consider Gaussian process state-space models (GP-SSMs), in which a GP is used to represent an unknown state transition function. A simple nonlinear example illustrates how a transition function learned by GP regression can be incorporated into a state-space model and how particle filtering and fixed-lag smoothing can be used for state estimation. The purpose is not to provide a comprehensive treatment of GP-SSMs, but rather to illustrate the connection between Gaussian process regression and nonlinear state-space modeling.

Through these examples, we aim to show how Gaussian processes provide a flexible and interpretable framework for time series modeling and how the choice and combination of kernel functions can be adapted to different types of temporal structure.

\section{Gaussian Processes for Time Series Modeling}
\subsection{Gaussian Processes and Kernel Functions}

A Gaussian process with mean function \(m(x)\) and covariance function \(k(x,x')\), denoted by
$$
f(x)\sim {\mathcal GP}\bigl(m(x),k(x,x')\bigr),
$$
is a collection of random variables \(\{f(x)\mid x\in\mathcal{X}\}\) such that, for any finite set of input points \(x_1,x_2,\ldots,x_n\), the corresponding random vector
$$
\left(f(x_1),f(x_2),\ldots,f(x_n)\right)^T
$$
follows a multivariate normal distribution. Its mean vector is
$$
\left(m(x_1),m(x_2),\ldots,m(x_n)\right)^T,
$$
and its covariance matrix \(K\) has \((i,j)\)-th element \(k(x_i,x_j)\). Thus, a Gaussian process defines a probability distribution over functions, with the covariance function \(k(x,x')\) specifying the dependence between function values at different input points.

In machine learning, the covariance function \(k(x,x')\) is commonly referred to as a kernel function. Some frequently used kernel functions are listed below.
\begin{itemize}
\item Gaussian (rbf, radial basis function) kernel: 
$
k(x,x')=\tau^2 \exp\left\{-\frac{(x-x')^2}{2\ell^2}\right\}.
$
\item Exponential (exp) kernel:
$
k(x,x')=\tau^2
\exp\left(-\frac{|x-x'|}{\ell}\right).
$
\item Linear kernel:
$
k(x,x')=\tau^2(x-c)^T(x'-c).
$
\item Periodic kernel:
$
k(x,x')=\tau^2
\exp\left\{
-\frac{1-\cos\left(2\pi(x-x')/p\right)}{\ell^2}
\right\}.
$
\item Rational quadratic (rq) kernel:
$
k(x,x')=\tau^2
\left\{
1+\frac{(x-x')^2}{2\alpha\ell^2}
\right\}^{-\alpha}.
$
\item Mat\'{e}rn kernel:
$
k(x,x')
=
\tau^2\frac{2^{1-\nu}}{\Gamma(\nu)}
\left(
\frac{\sqrt{2\nu}|x-x'|}{\ell}
\right)^\nu
K_\nu\left(
\frac{\sqrt{2\nu}|x-x'|}{\ell}
\right).
$
\end{itemize}
Here, \(\tau^2\) is the signal variance, \(\ell\) is the length scale, \(p\) is the period, \(\alpha\) is the shape parameter of the rational quadratic kernel, \(c\) is the offset or center of the linear kernel, and \(\nu\) is the smoothness parameter of the Mat\'{e}rn kernel. In the Mat\'{e}rn kernel, \(K_\nu(\cdot)\) denotes the modified Bessel function of the second kind.

For \(\nu=1/2,\,3/2\) and \(5/2\), the Mat\'{e}rn kernel has the following simple forms:
\begin{eqnarray}
  \left\{ \begin{array}{ll}
     k_{1/2}(x,x') &= \tau^2\exp\biggl(-\dfrac{|x-x'|}{\ell}\biggr), \qquad \textrm{(exponential kernel)}, \\[3mm]
     k_{3/2}(x,x') &= \tau^2 \biggl(1+\dfrac{\sqrt{3}|x-x'|}{\ell}\biggr)
                      \exp\biggl(-\dfrac{\sqrt{3}|x-x'|}{\ell}\biggr), \\[3mm]
     k_{5/2}(x,x') &= \tau^2 \biggl\{1+\dfrac{\sqrt{5}|x-x'|}{\ell} + \dfrac{5(x-x')^2}{3\ell^2} \biggr\}
                      \exp\biggl(-\dfrac{\sqrt{5}|x-x'|}{\ell}\biggr).
  \end{array} \right.
\end{eqnarray}
As \(\nu\rightarrow\infty\), the Mat\'{e}rn kernel converges to the rbf kernel,
$$
k_{\infty}(x,x') = \tau^2 \exp\left\{-\frac{(x-x')^2}{2\ell^2}\right\}.
$$

For a set of training inputs \(x_1,\ldots,x_n\), the kernel matrix
$$
K_{TT} = \Bigl\{k(x_i,x_j)\Bigr\}_{i,j=1}^{n}
$$
is a covariance matrix and is therefore symmetric and positive semidefinite. Kernel functions also have useful closure properties. In particular, if \(k_1(x,x')\) and \(k_2(x,x')\) are valid kernels, then both their sum,
$$
k(x,x') = k_1(x,x')+k_2(x,x'),
$$
and their product,
$$
k(x,x') = k_1(x,x')k_2(x,x'),
$$
are also valid kernels.

The kernel function plays a central role in Gaussian process modeling because it determines the dependence structure among function values and thereby controls the types of functions that are likely under the GP prior. By choosing an appropriate kernel, properties such as smoothness, periodicity, linear trends, and characteristic length scales can be incorporated into the prior distribution over functions. The observed data then update this prior information to produce the posterior and predictive distributions.

More flexible covariance structures can be constructed by combining basic kernels. For example, an additive kernel
$$
k(x,x')=k_1(x,x')+k_2(x,x')
$$
can be interpreted as representing the sum of independent latent GP components associated with the two kernels. Such a construction is useful, for example, for separating trend and periodic components. Product kernels,
$$
k(x,x')=k_1(x,x')k_2(x,x'),
$$
provide another way of combining kernel properties and can represent interactions between different forms of dependence. For example, multiplying an rbf kernel by a periodic kernel produces periodic dependence whose strength gradually decreases with increasing separation between the input points.

Thus, combinations of basic kernels can be used to represent complex behavior involving, for example,
\begin{itemize}
\setlength{\itemsep}{0pt}
\setlength{\topsep}{0pt}
\setlength{\parsep}{0pt}
\item trend and periodic variation,
\item long-term and local variation, and
\item linear and nonlinear behavior.
\end{itemize}
The choice and combination of kernel functions are therefore important aspects of Gaussian process modeling.

\begin{figure}[tbp]
\begin{center}
\includegraphics[width=140mm,angle=0,clip=]{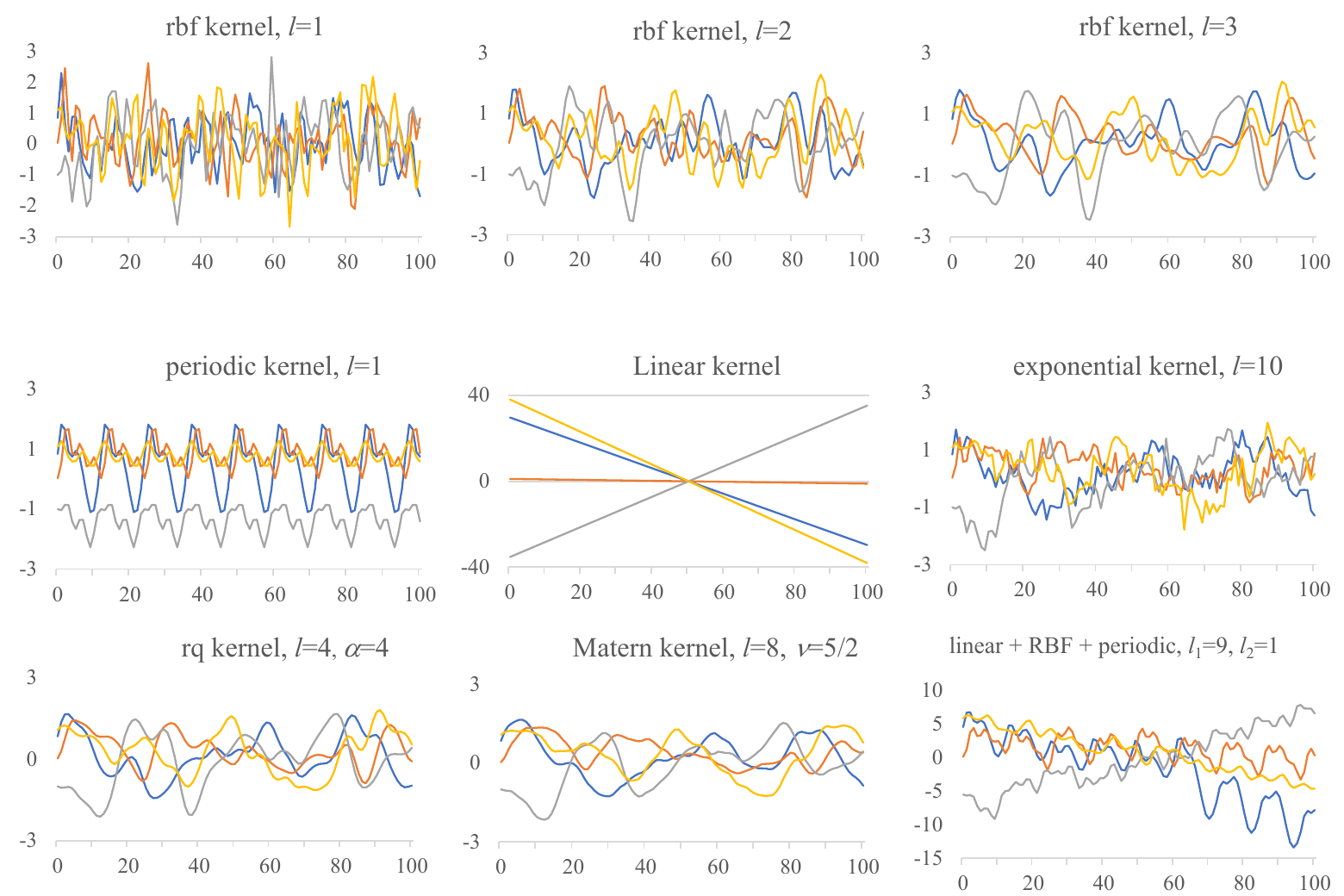}
\caption{Examples of Gaussian process realizations generated using various kernel functions. From top left to bottom right: rbf kernels with \(\ell=1,2,\) and \(3\); periodic kernel; linear kernel; exponential kernel; rq kernel with \(\alpha=\ell=4\); Mat'ern kernel with \(\ell=8\) and \(\nu=5/2\); and a composite linear+rbf+periodic kernel. Each panel shows four realizations.}
\label{Fig:kernel_functions}
\end{center}
\end{figure}

Figure \ref{Fig:kernel_functions} illustrates sample paths generated from Gaussian processes with several representative kernel functions. Each panel shows four realizations generated using the specified kernel. The first row shows realizations obtained using the rbf kernel for three different length scales, \(\ell=1,2\) and \(3\). As \(\ell\) increases, nearby function values remain strongly correlated over greater distances, resulting in smoother sample paths whose variations occur over longer input scales.

The second and third rows show realizations generated using the periodic, linear, exponential, rational quadratic, Mat\'{e}rn and composite linear+rbf+periodic kernels. The periodic kernel generates sample paths that repeat with the specified period. The exponential kernel produces less smooth sample paths than the rbf kernel, whereas the Mat\'{e}rn family provides intermediate degrees of smoothness controlled by the parameter \(\nu\). The rational quadratic kernel can represent dependence over a range of length scales. Finally, the composite kernel generates more complex sample paths by combining the characteristics of its constituent kernels.

\subsection{Gaussian Process Regression Model}

Suppose that a set of training data
$$
D=\{(x_1,y_1),\ldots,(x_n,y_n)\}
$$
is given, where, in the time series context, \(x_i\) denotes the observation time and \(y_i\) is the observed value of the time series at time \(x_i\). Gaussian process regression assumes the observation model
\begin{eqnarray}
y_i=f(x_i)+\varepsilon_i,\qquad
\varepsilon_i\sim{\mathcal N}(0,\sigma^2),
\end{eqnarray}
where the observation errors \(\varepsilon_i\) are mutually independent, and the latent function \(f(x)\) is assumed to follow a Gaussian process,
\begin{eqnarray}
f(x)\sim{\mathcal GP}\bigl(m(x),k(x,x')\bigr).
\end{eqnarray}
For simplicity, we assume \(m(x)=0\) in the following. A nonzero mean function can be accommodated by applying the same formulas to the centered observations.

Let
$$
y=(y_1,\ldots,y_n)^T
$$
denote the vector of training observations, and consider the prediction of observations
$$
y^*=(y_1^*,\ldots,y_m^*)^T
$$
at a set of prediction inputs
$$
x^*=(x_1^*,\ldots,x_m^*)^T.
$$
Define the kernel matrices for the training and prediction inputs as
\begin{eqnarray}
K_{TT} &=& \left[
   \begin{array}{ccc} k(x_1,x_1) &\cdots &k(x_1,x_n)\\
                      \vdots     &\ddots &\vdots    \\
                      k(x_n,x_1) &\cdots &k(x_n,x_n)
   \end{array} \right], \\[2mm]  
K_{TV} &=& \left[ \begin{array}{ccc}
           k(x_1,x_1^*) &\cdots &k(x_1,x_m^*) \\
           \vdots       &\ddots &\vdots       \\
           k(x_n,x_1^*) &\cdots &k(x_n,x_m^*)
   \end{array}\right], \\[2mm]
K_{VV} &=& \left[
   \begin{array}{ccc} k(x_1^*,x_1^*) &\cdots &k(x_1^*,x_m^*)\\
                      \vdots         &\ddots &\vdots        \\
                      k(x_m^*,x_1^*) &\cdots &k(x_m^*,x_m^*)
   \end{array}\right].
\end{eqnarray}
Here, \(K_{TT}\) is the covariance matrix for the latent function values at the training inputs, \(K_{TV}\) is the cross-covariance matrix between the training and prediction inputs, and \(K_{VV}\) is the covariance matrix for the latent function values at the prediction inputs.

Since the observations contain independent Gaussian noise, define the noise-augmented covariance matrices
\begin{eqnarray}
C_y=K_{TT}+\sigma^2I_n,
\qquad
C_{y^*}=K_{VV}+\sigma^2I_m.
\end{eqnarray}
Under the Gaussian process model, the joint distribution of the observed training values \(y\) and the future observations \(y^*\) is
\begin{eqnarray}
\left[ \begin{array}{c}  y \\[1mm]
                         y^*
       \end{array}\right]
\sim
{\mathcal N}\left( \left[ \begin{array}{c} 0 \\[1mm]
                                           0
                          \end{array}\right],
\left[
\begin{array}{cc}  C_y      & K_{TV} \\[1mm]
                   K_{TV}^T & C_{y^*}
\end{array} \right] \right).
\end{eqnarray}

It follows from the conditional distribution of a multivariate normal distribution that, given the training data \(D\), the predictive distribution of \(y^*\) is
\begin{eqnarray}
p(y^*\mid x^*,D) \sim  {\mathcal N}\bigl(\mu_V,\Sigma_V\bigr),
\end{eqnarray}
where
\begin{eqnarray}
\mu_V &=& K_{TV}^TC_y^{-1}y, \\[1mm]
\Sigma_V &=& C_{y^*}-K_{TV}^TC_y^{-1}K_{TV} \nonumber\\
         &=& K_{VV} + \sigma^2 I_m -K_{TV}^TC_y^{-1}K_{TV}.
\end{eqnarray}
Thus, the predictive mean is determined by the covariance between the prediction and training inputs, whereas the predictive covariance reflects both the uncertainty in the latent function and the observation noise.

If the objective is instead to predict the latent function values
$$
f^*=\bigl(f(x_1^*),\ldots,f(x_m^*)\bigr)^T,
$$
the corresponding posterior distribution is
\begin{eqnarray}
p(f^*\mid x^*,D) \sim {\mathcal N}\bigl(\mu_V,\Sigma_f\bigr),
\end{eqnarray}
where
\begin{eqnarray}
\Sigma_f = K_{VV}-K_{TV}^TC_y^{-1}K_{TV}.
\end{eqnarray}
Hence, the predictive covariance for future observations differs from that for the latent function by the observation-noise term:
$$
\Sigma_V = \Sigma_f + \sigma^2 I_m.
$$

\section{Fitting Gaussian Process Models to Time Series}

Let \(D\) denote the complete set of \(N\) time series observations. To evaluate both the fit to the observed data and the predictive performance of a Gaussian process model, we divide \(D\) into a training set \(D_T\) and an evaluation set \(D_V\):
$$
D=D_T\sqcup D_V,
$$
where \(D_T\) contains \(n\) observations and \(D_V\) contains \(m=N-n\) observations. Let
$$
y=(y_1,\ldots,y_n)^T
$$
and
$$
y^*=(y_1^*,\ldots,y_m^*)^T
$$
denote the observation vectors corresponding to \(D_T\) and \(D_V\), respectively. After arranging the training observations first and the evaluation observations second, the complete observation vector can be written as
\begin{eqnarray}
y^+ = \left[ \begin{array}{c} y  \\
                              y^*
             \end{array} \right].
\end{eqnarray}

Let \(\theta\) denote the vector of hyperparameters of the Gaussian process regression model, including the kernel parameters and the observation-noise variance when it is estimated jointly. The maximum likelihood estimate of \(\theta\) is obtained from the training data as
\begin{eqnarray}
   \hat{\theta} = \arg\max_{\theta}\log p(y\mid\theta),
\end{eqnarray}
where the marginal likelihood of the training observations is
\begin{eqnarray}
   p(y\mid\theta) = (2\pi)^{-n/2} |C_y(\theta)|^{-1/2}
                    \exp\left\{ -\frac{1}{2}y^TC_y(\theta)^{-1}y \right\},
   \label{Eq:marginal_likelihood}
\end{eqnarray}
with
\begin{eqnarray}
   C_y(\theta) = K_{TT}(\theta)+\sigma_n^2I_n.
\end{eqnarray}
Here, \(K_{TT}(\theta)\) is the \(n\times n\) kernel matrix evaluated at the training inputs, and \(\sigma_n^2\) denotes the observation-noise variance. For simplicity, a zero mean function is assumed throughout this section. If a nonzero mean function is used, the observation vector \(y\) in the above expressions is replaced by the corresponding residual vector.

\subsection{In-Sample Marginal Log-Likelihood}

Taking the logarithm of (\ref{Eq:marginal_likelihood}), the in-sample marginal log-likelihood evaluated at the maximum likelihood estimate \(\hat{\theta}\) is
\begin{align}
L_{\mathrm{in}} &= \log p(y\mid\hat{\theta}) \nonumber\\
                &= -\frac{1}{2} \left\{
                     y^TC_y(\hat{\theta})^{-1}y + \log|C_y(\hat{\theta})| + n\log(2\pi) \right\},
\end{align}
where
\begin{eqnarray}
   C_y(\hat{\theta}) = K_{TT}(\hat{\theta}) + \hat{\sigma}_n^2I_n.
\end{eqnarray}

The in-sample marginal log-likelihood measures how well the GP model, including its kernel structure and hyperparameters, accounts for the training data. The marginal likelihood itself incorporates a trade-off between goodness of fit and model flexibility through both the quadratic term and the determinant term. For comparison among kernel models having different numbers of estimated hyperparameters, we also consider the Akaike information criterion (AIC), defined by
\begin{eqnarray}
   \mathrm{AIC} = -2L_{\mathrm{in}} + 2d_\theta,
\end{eqnarray}
where \(d_\theta\) denotes the number of estimated hyperparameters. A smaller value of AIC indicates a better balance between fit and model complexity according to this criterion.

\subsection{Out-of-Sample Predictive Log-Likelihood}

To evaluate the predictive performance of a fitted Gaussian process model, we consider the conditional distribution of the evaluation data \(y^*\) given the training data \(y\). All kernel and covariance matrices in this subsection are evaluated at the hyperparameter estimate \(\hat{\theta}\) obtained solely from the training data.

The joint distribution of the training and evaluation observations is
\begin{eqnarray}
   \left[ \begin{array}{c} y \\
                           y^*
   \end{array} \right]
   \sim {\mathcal N} \left( \left[\begin{array}{c} 0 \\[2mm] 0 \end{array} \right], 
        \left[ \begin{array}{cc}
           C_y    & K_{TV}\\[1mm]
           K_{VT} & C_{y^*}
        \end{array} \right] \right),
\end{eqnarray}
where
$$
K_{VT}=K_{TV}^T,
$$
and
\begin{align}
   C_y     &= K_{TT}(\hat{\theta})+\hat{\sigma}_n^2I_n,\\
   C_{y^*} &= K_{VV}(\hat{\theta})+\hat{\sigma}_n^2I_m.
\end{align}
Here, \(K_{TV}\) is the \(n\times m\) cross-covariance matrix between the training and evaluation inputs, and \(K_{VV}\) is the \(m\times m\) kernel matrix evaluated at the evaluation inputs.

It follows from the conditional distribution of a multivariate normal distribution that
\begin{eqnarray}
   y^*\mid y,\hat{\theta} \sim {\mathcal N}\bigl(\mu_V,\Sigma_V\bigr),
\end{eqnarray}
where
\begin{align}
  \mu_V    &= K_{VT}C_y^{-1}y, \\[1mm]
  \Sigma_V &= C_{y^*}-K_{VT}C_y^{-1}K_{TV} \nonumber\\
           &= K_{VV}(\hat{\theta}) + \hat{\sigma}_n^2I_m - K_{VT}C_y^{-1}K_{TV}.
\end{align}

It is important to note that \(\Sigma_V\) is the predictive covariance matrix of the future observations \(y^*\). It therefore includes the observation-noise variance \(\hat{\sigma}_n^2I_m\). In contrast, the predictive covariance matrix of the corresponding latent function values would be
\begin{eqnarray}
   \Sigma_f = K_{VV}(\hat{\theta}) - K_{VT}C_y^{-1}K_{TV},
\end{eqnarray}
so that
$$
\Sigma_V = \Sigma_f+\hat{\sigma}_n^2I_m.
$$

The out-of-sample predictive log-likelihood is defined as the logarithm of the joint predictive density of the evaluation observations:
\begin{align}
   L_{\mathrm{out}} &= \log p(y^*\mid y,\hat{\theta}) \nonumber\\
                    &= -\frac{1}{2} \left\{ (y^*-\mu_V)^T \Sigma_V^{-1} (y^*-\mu_V)
                       + \log|\Sigma_V| + m\log(2\pi) \right\}.
\end{align}
This criterion evaluates predictive performance on observations that were not used to estimate the hyperparameters. A larger value of \(L_{\mathrm{out}}\) indicates a larger joint predictive density for the evaluation data, taking into account both the agreement of the predictive mean with the observations and the uncertainty represented by the predictive covariance matrix.

Finally, from the factorization
\begin{eqnarray}
   p(y,y^*\mid\hat{\theta}) = p(y\mid\hat{\theta})p(y^*\mid y,\hat{\theta}),
\end{eqnarray}
the log-likelihood of the complete data, evaluated at the hyperparameters estimated from the training data, can be written as
\begin{eqnarray}
   L_{\mathrm{all}} = L_{\mathrm{in}}+L_{\mathrm{out}}.
\end{eqnarray}
It should be emphasized that \(L_{\mathrm{all}}\) defined in this way is evaluated at \(\hat{\theta}\), which is estimated using only the training data. It is therefore generally different from the maximized marginal log-likelihood obtained by re-estimating the hyperparameters using all \(N\) observations.

\section{Examples}
\subsection{Ship Roll Motion Data}

\begin{figure}[h]
\begin{center}
\includegraphics[width=140mm,angle=0,clip=]{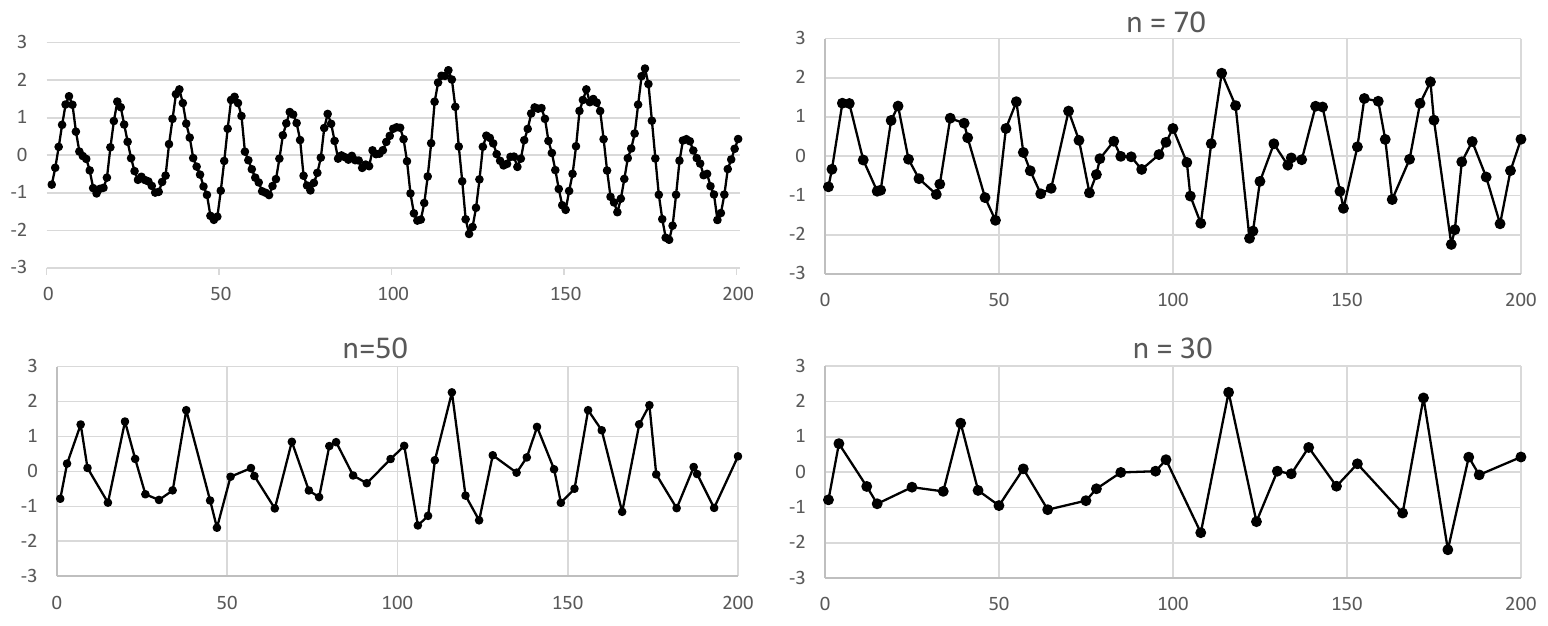}
\caption{Ship roll motion data and three subsamples used as training data.}
\label{Hakusan_data}
\end{center}
\end{figure}

We first consider the ship roll motion data as an example of a stationary time series. The upper-left panel of Figure \ref{Hakusan_data} shows the last \(N=200\) observations of the series \cite{Kitagawa 2021}.

To examine the performance of Gaussian process regression with different amounts of training data, three training datasets are constructed by selecting \(n=70\), \(50\) and \(30\) observations from the original series. The two endpoints, \(y(1)\) and \(y(N)\), are always included in each training dataset. The remaining \(n-2\) observations are selected by stratified random sampling. Specifically, the interval between the two endpoints is divided into \(n-2\) approximately equal subintervals, and one observation is randomly selected from each subinterval.

The other three panels of Figure \ref{Hakusan_data} show the training datasets obtained in this way. The datasets with \(n=70\) and \(n=50\) retain most of the characteristic features of the original time series, whereas the dataset with \(n=30\) loses some of the finer features.

\begin{table}[tbp]
\caption{Comparison of kernel models for the ship roll motion data. The table shows the in-sample marginal log-likelihood \(L_{\mathrm{in}}\), AIC, out-of-sample predictive log-likelihood \(L_{\mathrm{out}}\), and estimates of the kernel parameters and observation-noise variance. Here, \(n\) denotes the number of training observations. \(L_x=\mathrm{in}\) and \(L_x=\mathrm{out}\) indicate that the hyperparameters are selected by maximizing \(L_{\mathrm{in}}\) and \(L_{\mathrm{out}}\), respectively.}
\label{Tab:hakusan_GP-regression_MLE}
\begin{center}
\begin{small}
\begin{tabular}{c|c|c|ccr|cccccc} \hline
$n$ & $L_x$ & kernel & $L_{in}$ & AIC & $L_{out}\hspace{3mm}$ & $\tau_1^2$ & $\ell_1$ &$\tau_2^2$ & $\ell_2$ & $\alpha$ & $\sigma^2$ \rule[-5pt]{0pt}{16pt} \\
\hline
& in & rbf & $-73.850$ & 153.700 &  58.778 & 1.178 & 2.876 &   \rule[-2pt]{0pt}{14pt}&    &   & 0.0088 \\
& in & rq  & $-73.850$ & 155.700 &  58.778 & 1.178 & 2.876 &    &    & $4.9\times10^5$ & 0.0088 \\
70 & in & exp & $-89.882$ & 185.764 &$-69.530$& 0.955 & 2.879 &      &      &      & 0.0010 \\
& in &rbf+exp&$-73.850$& 157.700 &  59.112 & 1.178 & 2.876 & 0.007& 0.244&      & 0.0021 \\
\cline{2-12}
& out & rbf & $-82.482$ & 170.964 & $69.347$ & 0.537 & 2.211 &      &      &      & 0.0061 \\
& out &rbf+exp&$-82.482$ & 174.964 & $69.347$ & 0.537 & 2.211 & 0.004& 0.043&      & 0.0020 \\
\Xhline{0.8pt}
50 & in & rbf & $-65.315$ & 136.630 &$-98.808$& 1.175 & 2.862 &      &      & \rule[-2pt]{0pt}{12pt} & 0.0015 \\
\cline{2-12}
& out & rbf & $-67.462$ & 140.630 & $58.886$ & 0.706 & 2.311 &      &      &      & 0.0062 \\
\Xhline{0.8pt}
30 & in & rbf & $-42.326$ & 90.652 & $-78.889$& 0.991 & 1.708 &      &      & \rule[-2pt]{0pt}{12pt}& 0.0010 \\
\cline{2-12}
& out & rbf & $-42.795$ & 91.518 & $35.468$ & 0.862 & 2.366 & \rule[-4pt]{0pt}{14pt} &      &      & 0.0062  \\
\Xhline{1.2pt}
200 & in & rbf & $-7.220$ & 20.440 & ---\hspace{3mm}\rule[-2pt]{0pt}{14pt} & 0.934 & 2.380 &    &    &    & 0.0062 \\
\hline
\end{tabular}
\end{small}
\end{center}
\end{table}

Table \ref{Tab:hakusan_GP-regression_MLE} summarizes the Gaussian process models fitted to the three training datasets. The rbf kernel is used as the basic kernel. Since the number of observations used to evaluate \(L_{\mathrm{in}}\) changes with the training sample size \(n\), the values of \(L_{\mathrm{in}}\) and AIC should not be directly compared across different values of \(n\). Similarly, the rows obtained by optimizing different criteria, \(L_x=\mathrm{in}\) and \(L_x=\mathrm{out}\), have different interpretations and should be compared with care.

For \(n=70\), the rational quadratic (rq), exponential (exp), and additive rbf+exp kernels are also examined. For the rq kernel, the estimated value of \(\alpha\) is extremely large. Consequently, the estimates of the other parameters and the values of both \(L_{\mathrm{in}}\) and \(L_{\mathrm{out}}\) are virtually identical to those obtained with the rbf kernel. This is consistent with the fact that the rq kernel approaches the rbf kernel as \(\alpha\rightarrow\infty\). Because the rq model contains one additional hyperparameter, its AIC is slightly larger than that of the rbf model.

In contrast, the exponential kernel gives substantially smaller values of both \(L_{\mathrm{in}}\) and \(L_{\mathrm{out}}\) than the rbf kernel. The additive rbf+exp kernel gives essentially the same \(L_{\mathrm{in}}\) as the rbf kernel alone and only a very small increase in \(L_{\mathrm{out}}\). Moreover, the estimated signal variance of the exponential component is very small. These results indicate that adding the exponential component provides little improvement for these data.

The rows labeled \(L_x=\mathrm{out}\) show the results obtained by selecting the hyperparameters so as to maximize \(L_{\mathrm{out}}\). For the rbf kernel with \(n=70\), this gives a smaller signal variance, \(\tau^2=0.537\), and a shorter length scale, \(\ell=2.211\), than the corresponding estimates \(\tau^2=1.178\) and \(\ell=2.876\) obtained by maximizing \(L_{\mathrm{in}}\). The observation-noise variance is also slightly smaller. At the same time, \(L_{\mathrm{out}}\) increases from \(58.778\) to \(69.347\), at the expense of a decrease in \(L_{\mathrm{in}}\).

An even more pronounced difference is observed for \(n=50\) and \(n=30\). When the hyperparameters are estimated from the training data, the corresponding values of \(L_{\mathrm{out}}\) are \(-98.808\) and \(-78.889\), respectively. Maximization of \(L_{\mathrm{out}}\) increases these values to \(58.886\) and \(35.468\). These results illustrate that hyperparameters providing the best fit to the training data do not necessarily provide the best predictive distribution for the evaluation data.

It should be emphasized, however, that the evaluation observations are themselves used to select the hyperparameters in the rows labeled \(L_x=\mathrm{out}\). These rows therefore do not represent genuine out-of-sample prediction. Rather, they are included to illustrate the possible difference between hyperparameters selected according to in-sample fit and those selected according to predictive performance. In practical applications, a separate validation set or an appropriate cross-validation procedure would be required for such hyperparameter selection if an independent assessment of predictive performance were desired.

\begin{figure}[tbp]
\begin{center}
\includegraphics[width=140mm,angle=0,clip=]{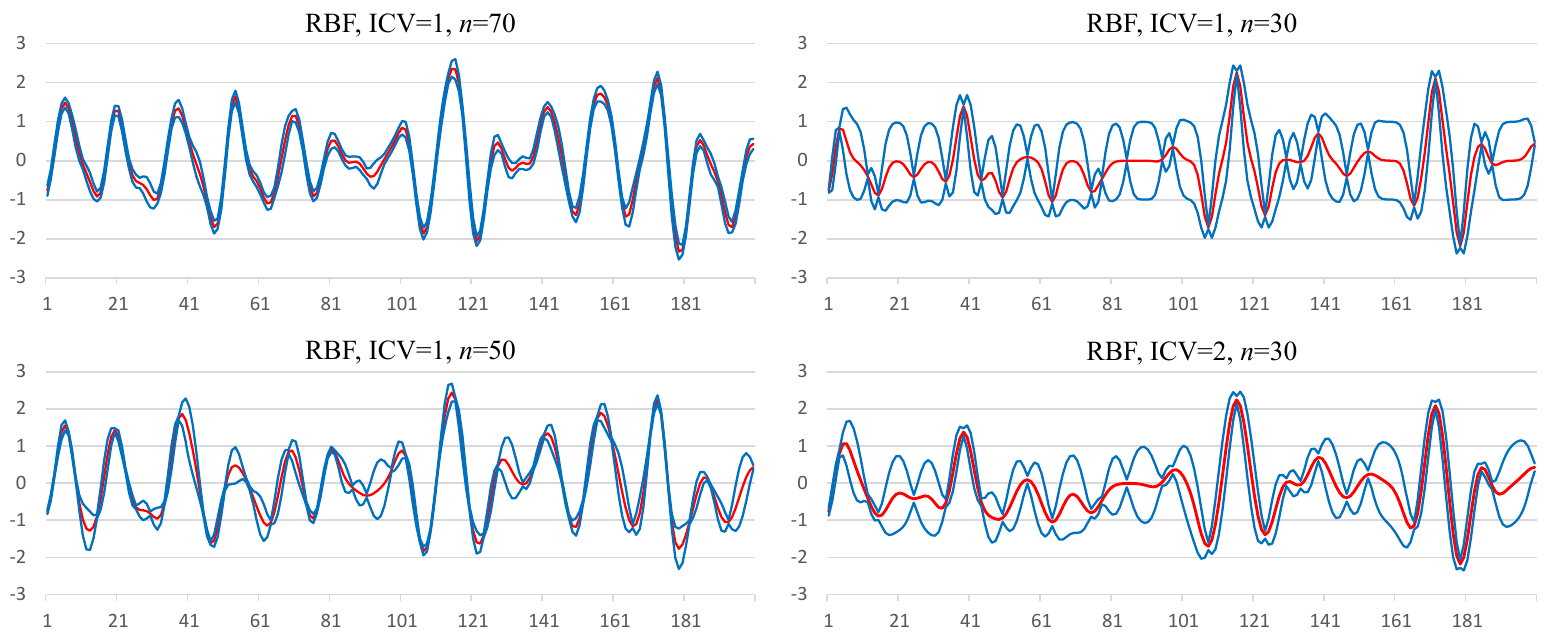}
\caption{Posterior distributions of observations of the ship roll motion data using the rbf kernel with \(n=70\), \(50\), and \(30\) training observations. Red curves show posterior means and the blue ones show show one-standard-deviation uncertainty band.}
\label{Fig:Hakusan_rbf}
\end{center}
\end{figure}

For \(n=50\) and \(n=30\), only the results obtained with the rbf kernel are presented. For reference, the last row of the table shows the result obtained by fitting the rbf kernel to all \(N=200\) observations.

Figure \ref{Fig:Hakusan_rbf} shows the posterior distributions obtained using the rbf kernel with hyperparameters estimated by maximizing \(L_{\mathrm{in}}\). For \(n=70\) and \(n=50\), although only 35\% and 25\% of the original observations, respectively, are used for training, the posterior means reproduce the main features of the original time series remarkably well. Thus, for these data, the smooth covariance structure represented by the rbf kernel allows the overall behavior of the series to be reconstructed from a relatively sparse set of observations.

For \(n=30\), corresponding to only 15\% of the original observations, the reconstruction becomes less accurate. This is consistent with the fact that the training dataset itself no longer retains some of the finer features of the original series, as can be seen in Figure \ref{Hakusan_data}. The large difference between the values of \(L_{\mathrm{out}}\) obtained by in-sample and out-of-sample optimization in Table \ref{Tab:hakusan_GP-regression_MLE} also indicates that, with such a sparse training sample, the predictive distribution is sensitive to the choice of hyperparameters.

\begin{figure}[tbp]
\begin{center}
\includegraphics[width=140mm,angle=0,clip=]{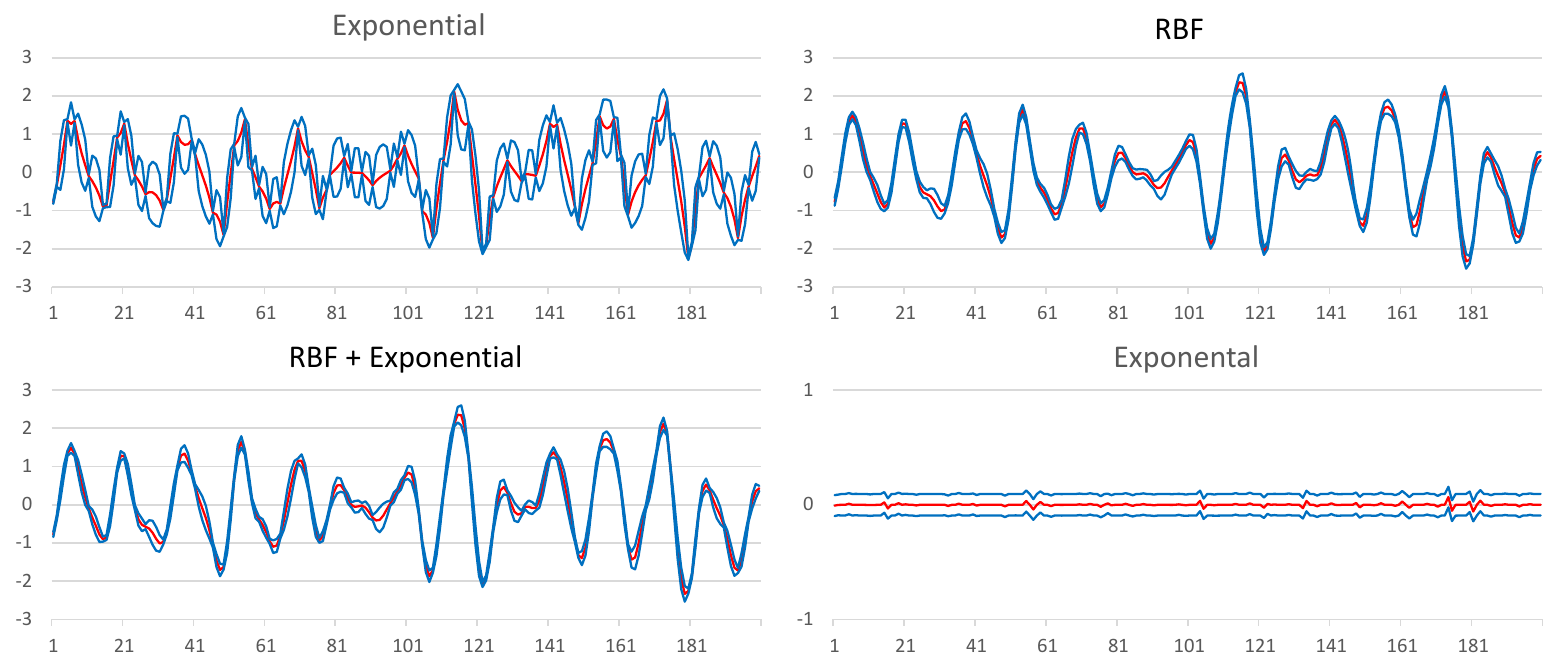}
\caption{Left colum shows the posterior distributions of observations of the ship roll motion data obtained using the exponential kernel (top) and additive rbf+exp kernel (bottom) for training data with \(n=70\). Right column shows the contribution of the rbf kernel (top) and exponential kernel (bottomn) in the additive rbf+exp kernel.}
\label{Fig:Hakusan_rbf-exponential}
\end{center}
\end{figure}

Left column of Figure \ref{Fig:Hakusan_rbf-exponential} compares the posterior distributions obtained using the exponential kernel (top) and the additive rbf+exp kernel (bottomn) for the \(n=70\) training data. With the exponential kernel, the posterior mean, shown in red, varies less smoothly than that obtained with the rbf kernel, and the one-standard-deviation uncertainty band, shown in blue, is substantially wider. This behavior is consistent with the lower smoothness implied by the exponential kernel.
With the additive rbf+exp kernel, a similar result as the rbf kernel is obtained.

For the additive rbf+exp kernel shown in the right panels, the posterior distribution can be decomposed into the contributions of the two latent GP components. The posterior mean of the rbf component is almost identical to that obtained with the rbf kernel alone. In contrast, the posterior mean of the exponential component remains close to zero, although its uncertainty band is relatively wide. Thus, most of the systematic variation in the series is represented by the rbf component, while the exponential component makes only a small additional contribution.

\subsection{Sunspot Number Data}

\begin{figure}[h]
\begin{center}
\includegraphics[width=140mm,angle=0,clip=]{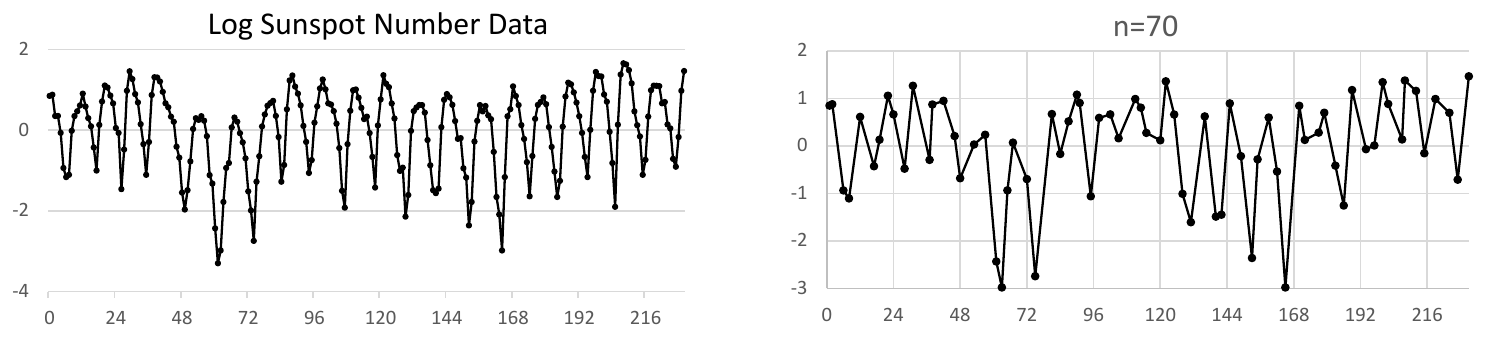}
\caption{Logarithm of annual sunspot number data (left) and training data obtained by stratified sampling with $n$=70.}
\label{Fig:Sunspot_data}
\end{center}
\end{figure}

As an example of a time series exhibiting quasi-periodic behavior, we consider the annual sunspot numbers \cite{Kitagawa 2021}. The left panel of Figure \ref{Fig:Sunspot_data} shows the logarithm of the annual sunspot numbers, standardized to have zero mean and unit variance. The right panel shows the training data consisting of \(n=70\) observations selected by stratified random sampling, using the same procedure as in the preceding example.

Table \ref{Tab:sunspot_GP-regression_MLE} summarizes the GP regression models fitted to these training data. Four kernel models are considered: the rbf kernel, an additive rbf+periodic kernel, and two product kernels formed by multiplying the rbf kernel by the cosine and periodic kernels, respectively. For the three kernels containing a periodic factor or component, the period \(p\) is treated as an unknown hyperparameter and estimated from the data rather than being restricted to an integer value.

\begin{table}[bp]
\caption{Comparison of four kernel models for the standardized log-sunspot data. The table shows the in-sample marginal log-likelihood \(L_{\mathrm{in}}\), AIC, out-of-sample predictive log-likelihood \(L_{\mathrm{out}}\), and estimates of the kernel parameters and observation-noise variance. The training sample consists of \(n=70\) observations selected by stratified random sampling.}
\label{Tab:sunspot_GP-regression_MLE}
\begin{center}
\begin{small}
\begin{tabular}{c|ccr|cccccc} \hline
kernel & $L_{in}$ & AIC & $L_{out}\hspace{3mm}$ & $\tau_1^2$ & $\ell_1$ &$\tau_2^2$ & $\ell_2$ & $p$ & $\sigma^2$ \rule[-5pt]{0pt}{16pt} \\ \hline
rbf           & $-97.907$ & 201.814 & $-86.123$ & 1.082 & 2.044 &   \rule[-2pt]{0pt}{14pt}&     &    & 0.08578 \\
rbf+periodic  & $-88.574$ & 189.148 & $-112.022$ & 0.516 & 2.226 & 0.414 & 0.538 & 11.262& 0.13691 \\
rbf$\times$cosine   & $-92.037$ & 192.075 &$-165.311$ & 0.574 &23.761 &      &      & 11.230 & 0.56849 \\
rbf$\times$periodic & $-84.980$ & 179.959 &$-114.256$ & 0.999 &20.070 &      & 1.074& 11.227 & 0.22912 \\
\hline
\end{tabular}
\end{small}
\end{center}
\end{table}

The rbf kernel gives the smallest in-sample marginal log-likelihood among the four models, but the largest out-of-sample predictive log-likelihood. In contrast, all three composite kernels improve both \(L_{\mathrm{in}}\) and AIC relative to the rbf kernel. Among them, the rbf\(\times\)periodic kernel gives the largest \(L_{\mathrm{in}}\) and the smallest AIC. This improvement in the fit to the training data, however, does not result in better prediction of the evaluation data. The rbf kernel alone gives the largest \(L_{\mathrm{out}}\), \(-86.123\). Among the three composite kernels, the additive rbf+periodic kernel gives the largest \(L_{\mathrm{out}}\), followed by the rbf\(\times\)periodic and rbf\(\times\)cosine kernels. Thus, for these data, a more elaborate representation of quasi-periodicity improves the fit to the training observations but does not necessarily improve out-of-sample predictive performance.

The estimated periods of the three models containing a periodic factor or component are \(11.262\), \(11.230\), and \(11.227\) years, respectively. The close agreement among these estimates indicates that the characteristic period of the sunspot series is consistently identified by the different kernel models.

\begin{figure}[h]
\begin{center}
\includegraphics[width=140mm,angle=0,clip=]{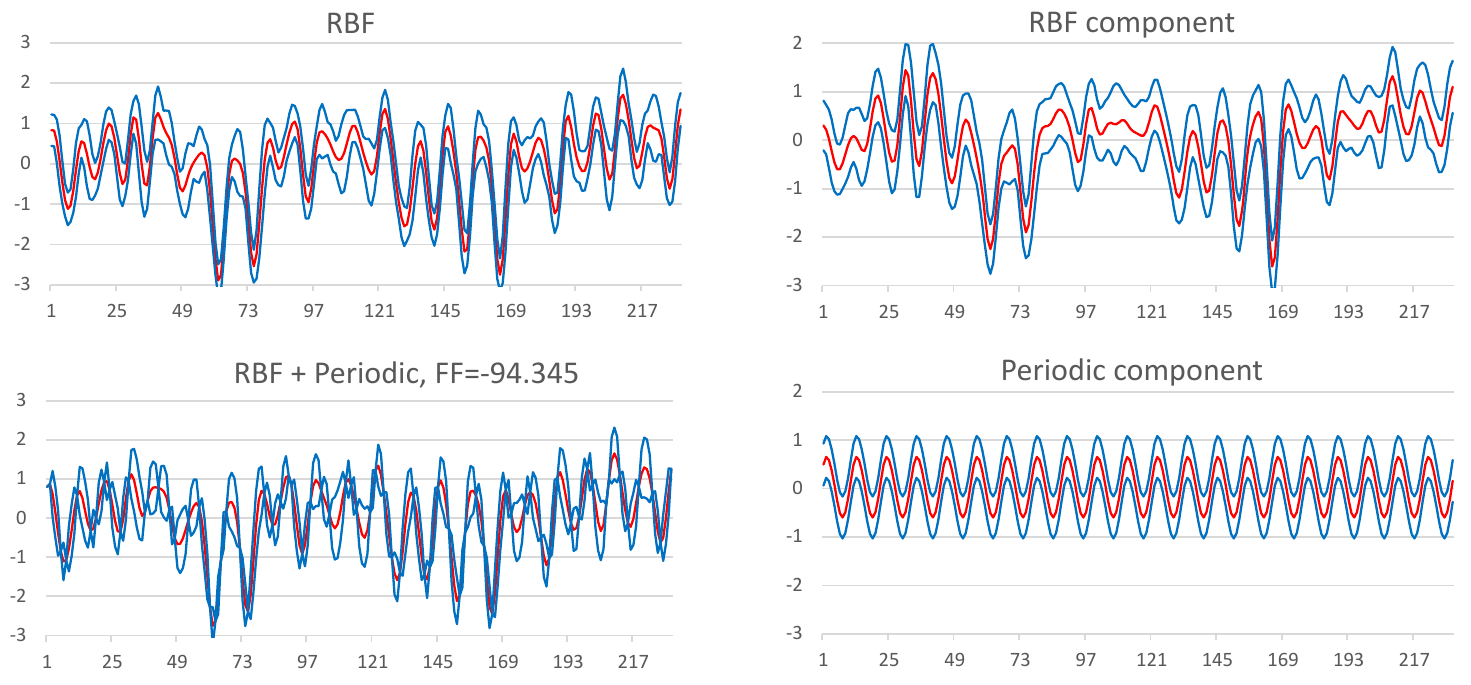}
\caption{Gaussian process modeling of the standardized log-sunspot data using the rbf and additive rbf+periodic kernels. Left column shows posterior distributions of observations by rbf kernel (top) and by addive rbf+periodic kernel (bottom). Right column show contribution of rbf kernel (top) and periodic kernel (bottom).}
\label{Fig:Sunspot_rbf_and_rbf-periodic}
\end{center}
\end{figure}

The top-left panel of Figure \ref{Fig:Sunspot_rbf_and_rbf-periodic} shows the predictive distribution obtained using the rbf kernel. Despite the absence of an explicit periodic component, the posterior mean closely follows the observed time series shown in Figure \ref{Fig:Sunspot_data}. Thus, the smooth and flexible covariance structure of the rbf kernel alone is able to reproduce much of the quasi-periodic behavior present in the sunspot data.

Other three panels show the predictive distribution obtained using the additive rbf+periodic kernel and the corresponding rbf and periodic components. For an additive kernel, these components have a direct probabilistic interpretation as latent Gaussian processes whose sum forms the overall process. The estimated periodic component has a period of \(p=11.262\) years. While this component represents strictly periodic variation, the rbf component accounts for longer-term and local departures from this regular pattern. Their sum therefore provides a flexible representation of the quasi-periodic behavior of the observed sunspot series.

\begin{figure}[tbp]
\begin{center}
\includegraphics[width=140mm,angle=0,clip=]{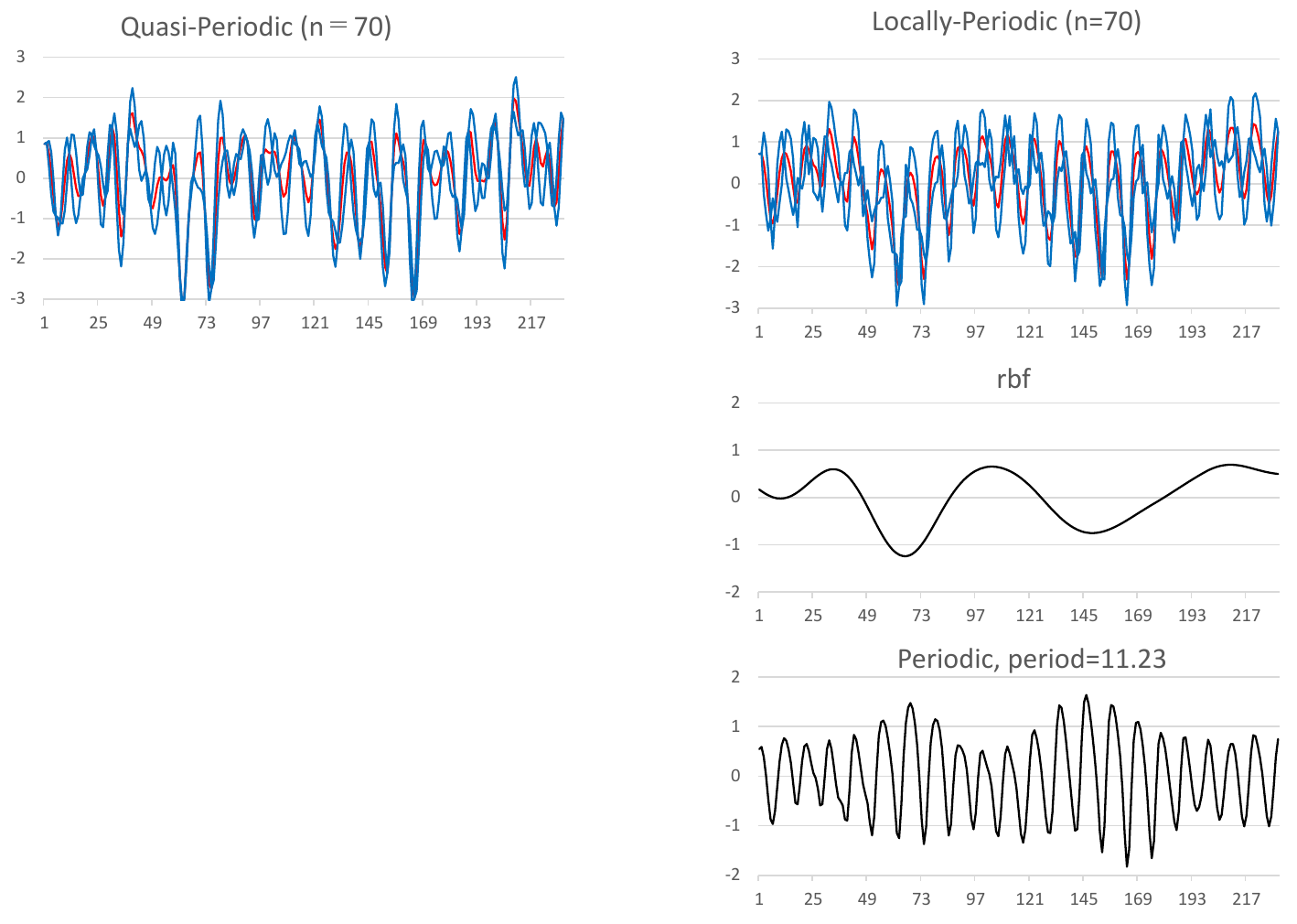}
\caption{Gaussian process modeling of the standardized log-sunspot data using product kernels. The left panel shows the result obtained with the rbf\(\times\)cosine quasi-periodic kernel. The right panels show the result obtained with the rbf\(\times\)periodic locally periodic kernel, together with representations of its rbf and periodic factors.}
\label{Fig:Sunspot_product_GP}
\end{center}
\end{figure}

Figure \ref{Fig:Sunspot_product_GP} compares the predictive distributions obtained using the two product kernels. The left panel shows the result for the rbf\(\times\)cosine kernel, which we refer to here as a quasi-periodic kernel. The right panels show the predictive distribution obtained using the rbf\(\times\)periodic kernel, referred to here as a locally periodic kernel, together with representations of its rbf and periodic factors.

Unlike the components of an additive kernel, the factors of a product kernel do not, in general, provide a decomposition into latent Gaussian processes. To see this, suppose that
$$
f_1\sim\mathcal{GP}(0,k_1) \qquad\textrm{and}\qquad f_2\sim\mathcal{GP}(0,k_2)
$$
are independent Gaussian processes. The product \(f_1f_2\) has mean zero and covariance function \(k_1k_2\), which are the same first two moments as those of a Gaussian process
$$
g\sim\mathcal{GP}(0,k_1k_2).
$$
However, \(f_1f_2\) is generally not a Gaussian process and is therefore not identical to \(g\).

Accordingly, the rbf and periodic factors shown in the middle-right and lower-right panels are constructed so that their product reproduces the posterior mean shown by the red curve in the upper-right panel. These factor representations are useful for visualizing how the two kernels contribute to the shape of the fitted function, but they should not be interpreted as a probabilistic decomposition of the posterior process. In particular, the posterior distribution under the product kernel cannot, in general, be obtained by multiplying posterior distributions associated with the individual kernels.

\begin{figure}[tbp]
\begin{center}
\includegraphics[width=120mm,angle=0,clip=]{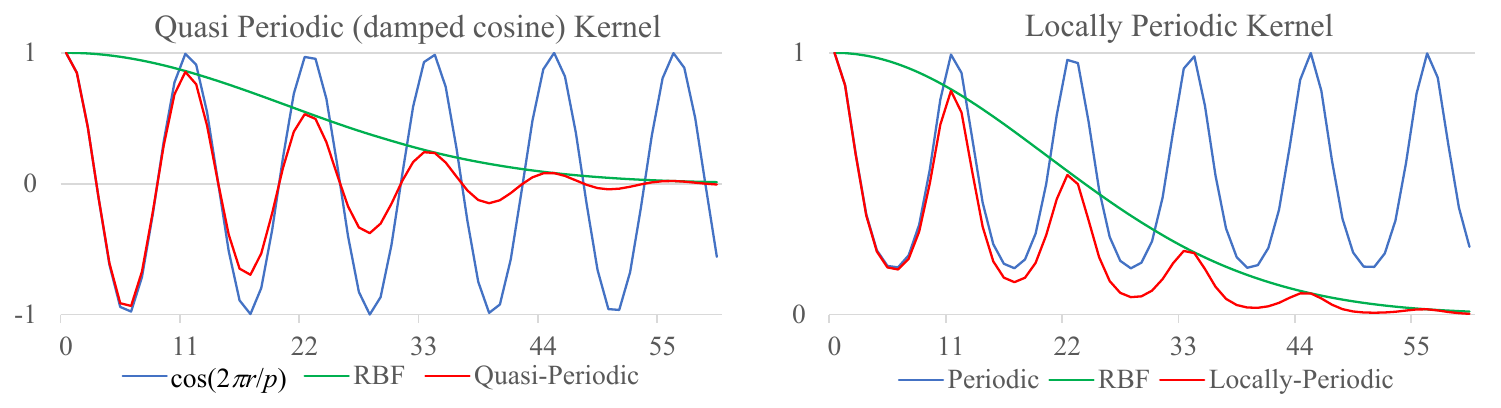}
\caption{Kernel functions for the two product kernels used for the sunspot data: the rbf\(\times\)cosine quasi-periodic kernel and the rbf\(\times\)periodic locally periodic kernel. The constituent kernel functions and their products are shown to illustrate the effects of the individual kernel factors.}
\label{Fig:Sunspot_product_kernels}
\end{center}
\end{figure}

The characteristics of the two product kernels can be understood more directly by examining the kernel functions themselves. Figure \ref{Fig:Sunspot_product_kernels} displays the constituent kernels and their products.
The left panels show the rbf\(\times\)cosine quasi-periodic kernel. If \(h=x-x'\) denotes the lag, this kernel has the form
$$
k_{\mathrm{QP}}(h) = \tau^2 \exp\left(-\frac{h^2}{2\ell^2}\right)\cos\left(\frac{2\pi h}{p}\right).
$$
It is a damped oscillatory covariance function whose amplitude decreases as \(|h|\) increases. The cosine factor produces alternating positive and negative dependence, while the rbf factor controls the rate at which the magnitude of this dependence decreases with increasing lag.

The right panels show the rbf\(\times\)periodic locally periodic kernel. Apart from the parameterization of the signal variance, it has the form
$$
k_{\mathrm{LP}}(h) = \tau^2 \exp\left\{ -\frac{h^2}{2\ell_1^2} -\frac{1-\cos(2\pi h/p)}{\ell_2^2} \right\}.
$$
This kernel is positive and exhibits peaks at integer multiples of the period \(p\). The heights of these peaks decrease with increasing \(|h|\) because of the rbf factor. Thus, observations separated by approximately one or more periods can remain strongly related when they are relatively close in time, while this periodic similarity gradually weakens over longer time intervals.

\subsection{Seasonal Data}

\begin{figure}[h]
\begin{center}
\includegraphics[width=140mm,angle=0,clip=]{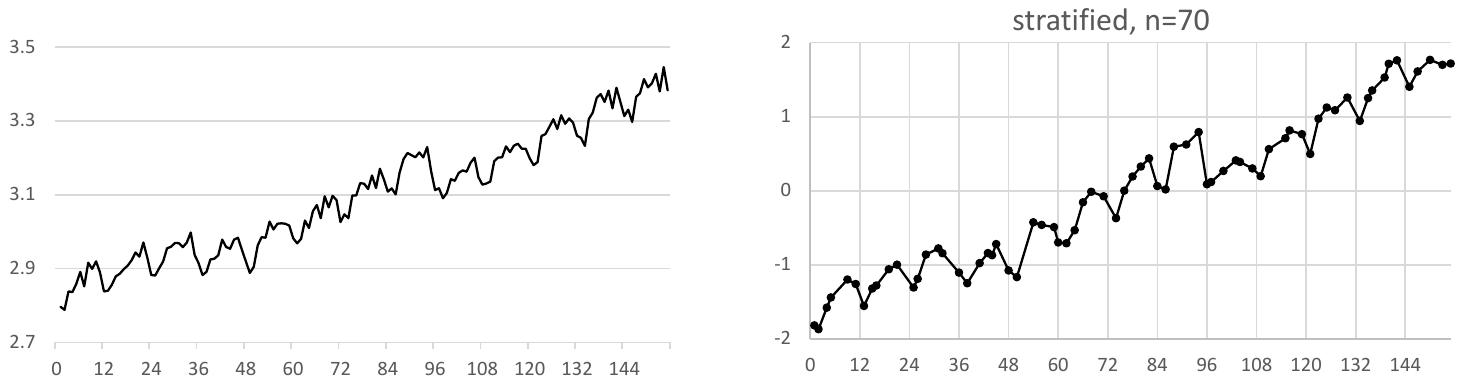}
\caption{Log-transformed Whard data and the training data with \(n=70\) observations selected by stratified random sampling.}
\label{whard_data}
\end{center}
\end{figure}

As a third example, we consider a time series exhibiting both trend and seasonal variation. Decomposition of a time series into trend and seasonal components has also been extensively studied using state-space modeling \cite{KG 1984,KG 1996}.
The left panel of Figure \ref{whard_data} shows the logarithm of the wholesale hardware (Whard) data, and the right panel shows the training dataset consisting of \(n=70\) observations selected by the same stratified random sampling procedure as in the preceding examples \cite{Kitagawa 2021}.

To represent the different types of variation in this time series, we consider an additive composite kernel consisting of linear, rbf and periodic kernels,
$$
k(x,x')
=
k_L(x,x')+k_R(x,x')+k_P(x,x').
$$
The three kernels are intended to represent the long-term trend, relatively short-term smooth variation, and regular seasonal variation, respectively. Because the kernel is additive, the corresponding GP can be interpreted as the sum of three independent latent Gaussian processes associated with these three sources of variation.

Table \ref{Tab:whard_GP-regression_MLE} summarizes the parameter estimates and model comparison criteria for GP regression models fitted to the training dataset with \(n=70\). The first row corresponds to the full model containing all three kernel components. Among the models considered, the full model gives the largest in-sample marginal log-likelihood \(L_{\mathrm{in}}\), the smallest AIC, and the largest out-of-sample predictive log-likelihood \(L_{\mathrm{out}}\).

\begin{table}[tbp]
\caption{Comparison of kernel models for the seasonal Whard data. The training sample consists of \(n=70\) observations. The table shows the in-sample marginal log-likelihood \(L_{\mathrm{in}}\), AIC, out-of-sample predictive log-likelihood \(L_{\mathrm{out}}\), and estimates of the kernel parameters and observation-noise variance. \(L=1\), \(R=1\), and \(P=1\) indicate that the linear, rbf, and periodic kernels, respectively, are included in the additive composite kernel. The last row shows the results obtained using all \(N=155\) observations.}
\label{Tab:whard_GP-regression_MLE}
\begin{center}
\medskip
\begin{small}\tabcolsep=1.5mm
\begin{tabular}{ccc|rrr|cccccc} \hline
L & R & P & $L_{in}$ & AIC & $L_{out}\hspace{3mm}$ & $\tau^2_L$ & $\tau^2_R$ & $\tau^2_P$ & $\ell_R$ & $\ell_P$ & $\sigma^2$ \rule[-5pt]{0pt}{16pt} \\ \hline
1 & 1 & 1 & $44.422$ & $-76.844$ & $72.644$ & 0.00045 & 0.02291 & 0.00137 & 7.772 & 0.499 & 0.00481 \rule[-2pt]{0pt}{14pt} \\
\hline
0 & 1 & 1 & $30.508$ & $-51.015$ & $66.810$ & ---\rule[-2pt]{0pt}{14pt} & 0.72597 & 0.00210 & 16.599 & 0.523 & 0.00632 \\
1 & 0 & 1 & $14.427$ & $-20.853$ & $19.997$ & 0.00047 & --- & 0.00781 & --- & 0.664 & 0.02747 \\
1 & 1 & 0 & $13.897$ & $-19.795$ & $-24.850$ & 0.00048 & 0.05341 & --- & 1.869 & --- & 0.00275 \\
\Xhline{0.4pt}
0 & 0 & 1 & $-101.029$ & 208.058 & $-118.996$ & --- & --- & $8.2!\times!!10^{-11}$ & ---\rule[-2pt]{0pt}{14pt} & 1.523 & 1.04988 \\
0 & 1 & 0 & $-2.841$ & 11.682 & $6.542$ & --- & 1.36451 & --- & 0.473 & --- & 0.03878 \\
1 & 0 & 0 & $-1.045$ & 6.090 & $-7.455$ & 0.00048 & --- & --- & --- & --- & 0.05446 \\
\Xhline{1.0pt}
1 & 1 & 1 & $120.326$ & $-234.651$ & ---\hspace{3mm}\rule[-2pt]{0pt}{12pt} & 0.00046 & 0.02259 & 0.00087 & 6.578 & 0.469 & 0.07744 \\
\hline
\end{tabular}
\end{small}
\end{center}
\end{table}

Rows 2--4 of Table \ref{Tab:whard_GP-regression_MLE} show the results obtained by removing one kernel component at a time from the full model. Among these reduced models, the model without the linear kernel performs best according to all three criteria. Its out-of-sample predictive log-likelihood, \(L_{\mathrm{out}}=66.810\), is only moderately smaller than that of the full model, \(L_{\mathrm{out}}=72.644\). This suggests that the flexible rbf component can partially compensate for the absence of the linear component by representing both the long-term trend and smoother local variation.

The models obtained by excluding the rbf and periodic kernels have similar in-sample marginal log-likelihoods, \(14.427\) and \(13.897\), respectively, but their out-of-sample predictive performances differ substantially. The model without the rbf kernel gives \(L_{\mathrm{out}}=19.997\), whereas the model without the periodic kernel gives \(L_{\mathrm{out}}=-24.850\). Thus, although the two models fit the training data to a similar degree, explicitly including the periodic component is particularly important for predicting the seasonal variation in the evaluation data.

Rows 5--7 show the results obtained using each kernel separately. Among the single-kernel models, the linear kernel gives the largest \(L_{\mathrm{in}}\) and the smallest AIC, whereas the rbf kernel gives the largest \(L_{\mathrm{out}}\). The periodic kernel alone performs very poorly according to all three criteria. Its estimated signal variance is nearly zero, while the estimated observation-noise variance is very large. Thus, a strictly periodic covariance structure alone cannot account for the substantial trend and nonperiodic variation present in this time series.

For reference, the last row shows the result obtained by fitting the full three-component model to all \(N=155\) observations. Most of the estimated kernel parameters are reasonably close to those obtained from the \(n=70\) training observations. In particular, the estimated linear and rbf signal variances are very similar. The periodic signal variance decreases from \(0.00137\) to \(0.00087\), and the corresponding length scale changes only slightly, from \(0.499\) to \(0.469\). On the other hand, the estimated observation-noise variance increases substantially, from \(0.00481\) to \(0.07744\).

\begin{figure}[tbp]
\begin{center}
\includegraphics[width=140mm,angle=0,clip=]{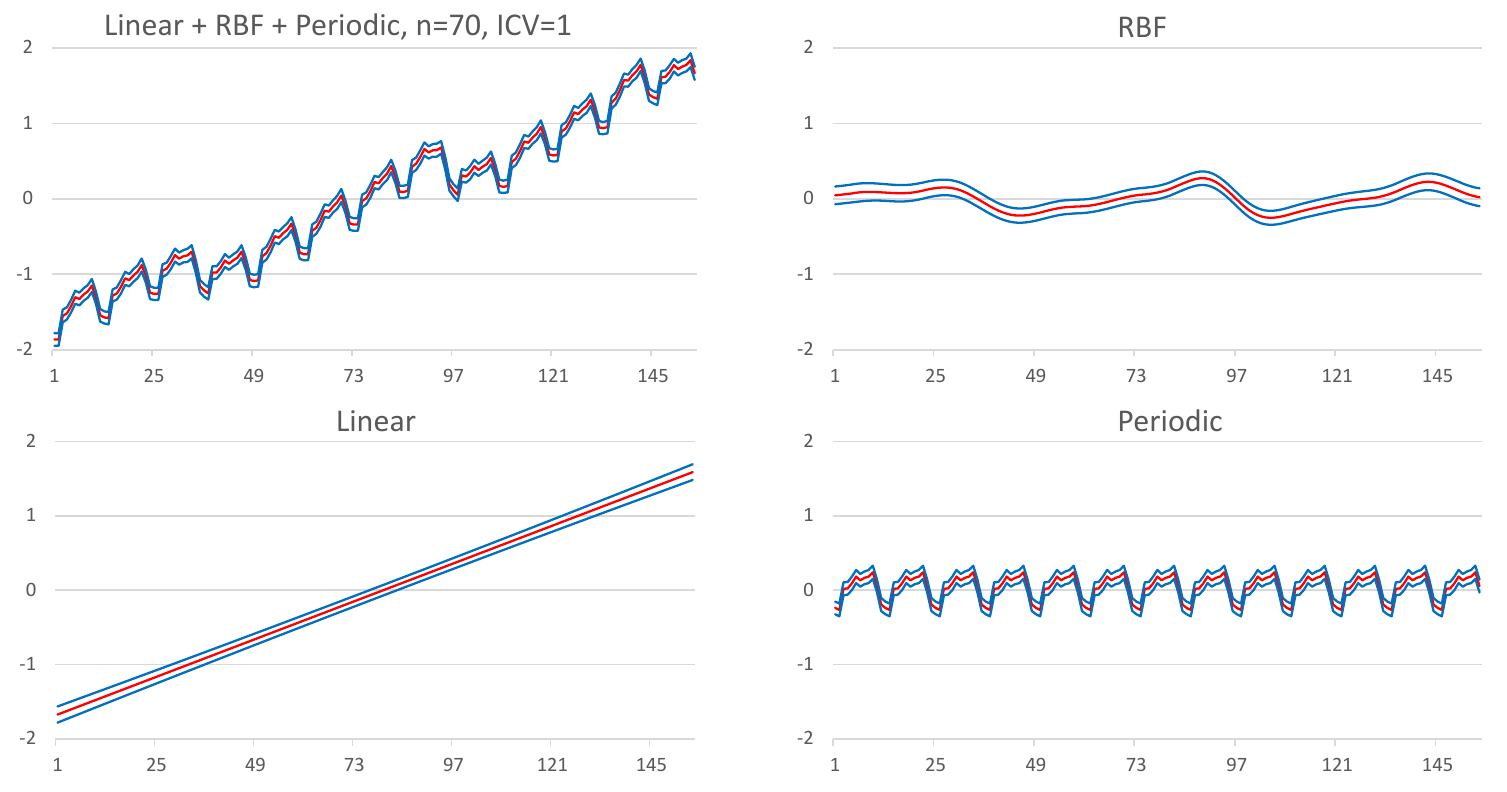}
\caption{Gaussian process modeling of the Whard data using the additive linear+rbf+periodic kernel. The predictive distribution of the time series and the posterior distributions of the three latent components are shown.}
\label{Fig:whard_LBP-kernel}
\end{center}
\end{figure}

Figure \ref{Fig:whard_LBP-kernel} shows the results obtained using the full additive kernel consisting of the linear, rbf, and periodic components. The upper-left panel shows the predictive distribution of the observed time series. The posterior mean captures both the overall trend and the seasonal variation well.

The other three panels show the posterior distributions of the linear, rbf, and periodic latent components, respectively. Because the composite kernel is additive, these components have a direct probabilistic interpretation as latent Gaussian processes whose sum forms the underlying process. The linear component represents the overall increasing trend, the rbf component captures relatively short-term smooth variation around the trend, and the periodic component represents the regular seasonal variation. Thus, the additive kernel provides an interpretable decomposition of the different sources of variation in the time series while also giving the smallest AIC and the largest \(L_{\mathrm{out}}\) among the models considered.

\begin{figure}[tbp]
\begin{center}
\includegraphics[width=140mm,angle=0,clip=]{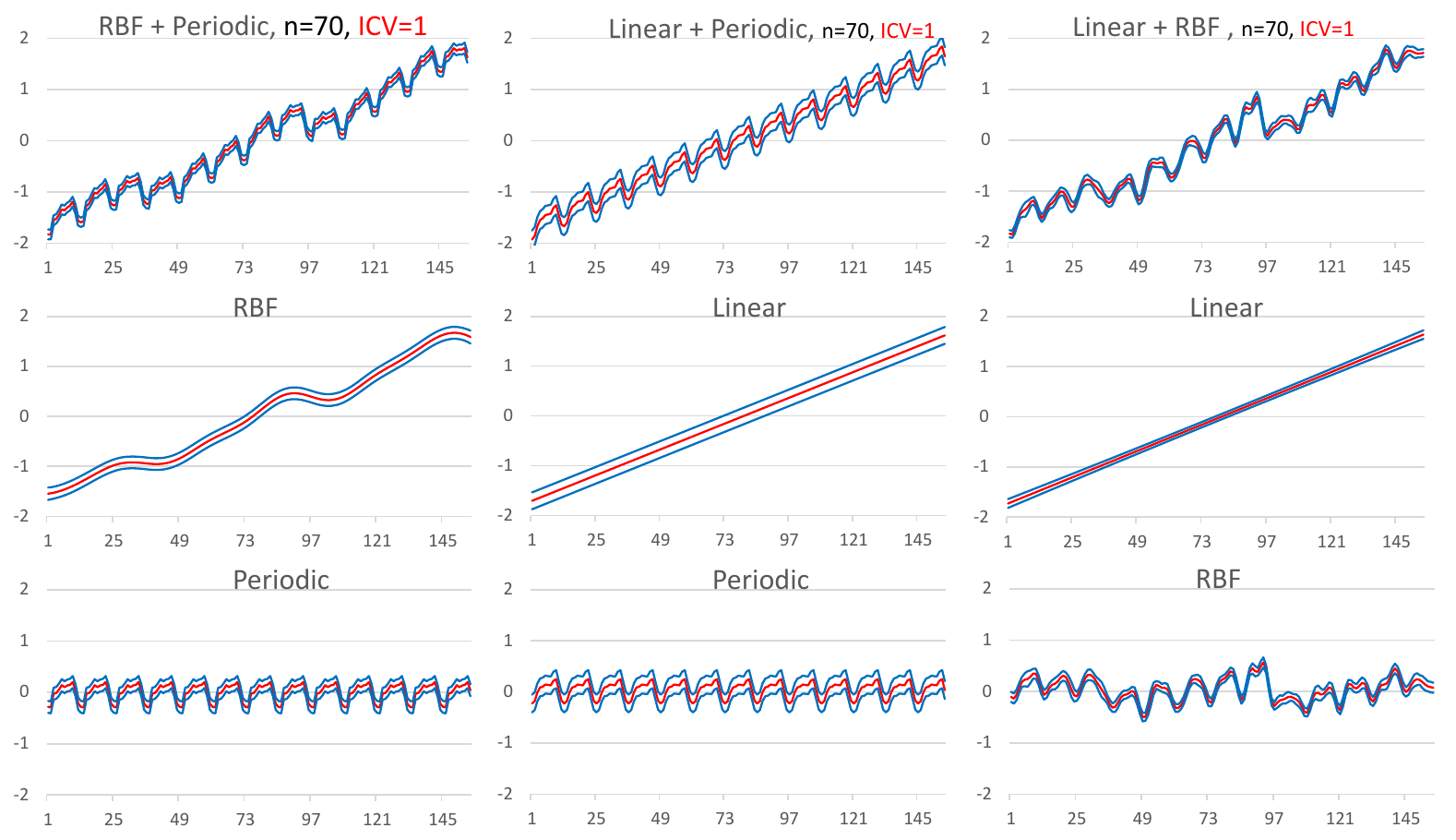}
\caption{Gaussian process modeling of the Whard data using two-component additive kernels. From left to right, the linear, rbf, and periodic components are excluded from the full model, respectively.}
\label{Fig:whard_2components-kernel}
\end{center}
\end{figure}

Figure \ref{Fig:whard_2components-kernel} shows the results obtained by excluding the linear, rbf, and periodic kernels from the full model, respectively, from left to right. The first row shows the predictive distributions of the observed time series, while the second and third rows show the posterior distributions of the two latent components included in each model.

When the linear kernel is excluded, the predictive distribution remains satisfactory because the flexible rbf component can represent both the long-term trend and local smooth variation. This is consistent with the relatively good values of \(L_{\mathrm{in}}\), AIC, and \(L_{\mathrm{out}}\) for the rbf+periodic model shown in Table \ref{Tab:whard_GP-regression_MLE}.

When the rbf kernel is excluded, the linear component represents the overall trend and the periodic component reproduces the seasonal variation, but local departures from these two systematic components cannot be represented adequately. When the periodic kernel is excluded, the rbf component attempts to represent both local and seasonal variation. Although the flexibility of the rbf kernel allows it to reproduce some of the oscillatory behavior, the regular seasonal pattern is considerably distorted. These results illustrate that a flexible kernel can partially compensate for a missing component, but such compensation may mix different sources of variation and make their interpretation less clear.

\begin{figure}[tbp]
\begin{center}
\includegraphics[width=140mm,angle=0,clip=]{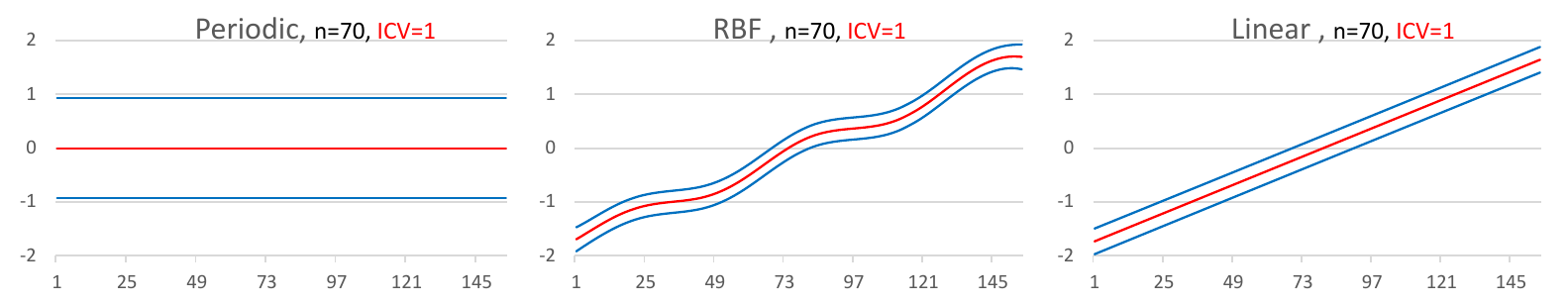}
\caption{Gaussian process modeling of the Whard data using the linear, rbf, and periodic kernels separately.}
\label{Fig:whard_1component-kernel}
\end{center}
\end{figure}

Figure \ref{Fig:whard_1component-kernel} shows the results obtained using each of the three kernels separately. The periodic kernel alone fails to capture the major characteristics of the time series. Its estimated signal variance is nearly zero, \(8.2\times10^{-11}\), while the observation-noise variance is estimated to be \(1.04988\). Thus, the fitted model attributes almost all of the variation to observation noise rather than to a periodic signal. The very small marginal log-likelihood and large AIC are consistent with this behavior.

The linear and rbf kernels provide substantially better representations than the periodic kernel alone. The linear kernel captures the overall trend but cannot explicitly represent local or seasonal fluctuations. The rbf kernel is more flexible and can represent both the long-term trend and some of the shorter-term variation. This flexibility is reflected in its larger out-of-sample predictive log-likelihood, \(6.542\), compared with \(-7.455\) for the linear kernel, even though the linear kernel gives a somewhat larger in-sample marginal log-likelihood and a smaller AIC.

Overall, this example illustrates the usefulness of an additive composite kernel for a time series containing several distinct forms of variation. For the Whard data, the linear, rbf, and periodic components play complementary roles: the linear kernel represents the long-term trend, the rbf kernel captures relatively short-term smooth variation around the trend, and the periodic kernel explicitly represents the seasonal structure. Although the flexibility of the rbf kernel allows it to compensate partially for the absence of some components, such compensation tends to mix different sources of variation. By explicitly combining all three kernels, the full model provides an interpretable decomposition of the time series and, among the models considered here, gives both the smallest AIC and the largest out-of-sample predictive log-likelihood.


\section{Gaussian Process State-Space Model}

In the preceding sections, Gaussian process regression was applied directly to observed time series. Gaussian processes can also be incorporated into state-space models by using a GP to represent an unknown state transition function. Models constructed in this way are commonly referred to as Gaussian process state-space models (GP-SSMs) \cite{ENDH 2017,FLSR 2013,FCR 2014}.

A simple GP-SSM can be written as
\begin{align}
x_n &= f(x_{n-1})+v_n, \qquad v_n\sim{\mathcal N}(0,\tau^2), \nonumber\\
y_n &= g(x_n)+w_n, \hspace{11mm} w_n\sim{\mathcal N}(0,\sigma^2),
\label{Eq:GP-SSM}
\end{align}
where \(x_n\) is the latent state, \(y_n\) is the observation, and \(v_n\) and \(w_n\) denote the process and observation noise, respectively. In a GP-SSM, the transition function \(f(\cdot)\) is modeled as a Gaussian process,
\begin{align}
f(x)\sim{\mathcal GP}\bigl(m(x),k(x,x')\bigr).
\end{align}
The GP therefore provides a flexible, nonparametric representation of the state transition function without requiring a specific parametric form for the state dynamics.

For example, if the observation function is \(g(x_n)=x_n\), the model reduces to
\begin{align}
x_n &= f(x_{n-1})+v_n, \nonumber\\
y_n &= x_n+w_n.
\end{align}
In this case, the GP represents the relationship between the previous state \(x_{n-1}\) and the current state \(x_n\), while the observations provide noisy measurements of the latent states. More general, possibly nonlinear, observation functions \(g(\cdot)\) can also be considered.

Inference in a GP-SSM generally involves two related tasks: learning the unknown transition function and its hyperparameters, and estimating the latent states from the observations. If the GP transition model has already been learned from suitable training data, it can be regarded as a probabilistic state transition model and incorporated directly into filtering or smoothing procedures.

For a fixed value of the previous state \(x_{n-1}\), the GP predictive distribution of the transition function is Gaussian. However, \(x_{n-1}\) itself is generally uncertain. Propagating this uncertainty through a nonlinear transition function and integrating over the distribution of \(x_{n-1}\) generally results in non-Gaussian filtering and smoothing distributions. Consequently, exact inference is usually not available, and various approximate inference methods have been developed for GP-SSMs, including particle-based and other approximate Bayesian methods.

In this book, we do not discuss the general theory or learning algorithms for GP-SSMs in detail. Instead, we briefly illustrate how a transition function learned by GP regression can be incorporated into a state-space model and how particle filtering can then be used to estimate the latent states sequentially from noisy observations.

\subsection{Particle Filter for GP-SSM}

Consider the GP-SSM in (\ref{Eq:GP-SSM}) and suppose that the GP transition model has already been learned from training data \(D\). Given the observations $ Y_{1:n}=\{y_1,\ldots,y_n\},$
the filtering problem is to estimate the filtering distribution $ p(x_n\mid Y_{1:n})$.
Suppose that, for a given previous state \(x_{n-1}\), GP regression provides the posterior distribution of the transition function $f(x_{n-1})\mid D \sim {\mathcal N}\bigl( \mu_f(x_{n-1}),s_f^2(x_{n-1})\bigr)$.

If the state equation is
$$
x_n=f(x_{n-1})+v_n, \qquad v_n\sim{\mathcal N}(0,q),
$$
the corresponding predictive distribution of the next state is
\begin{align}
p(x_n\mid x_{n-1},D) = {\mathcal N}\left( x_n\mid \mu_f(x_{n-1}),s_f^2(x_{n-1})+q \right).
\label{Eq:GP-transition}
\end{align}
Thus, the uncertainty in the GP estimate of the transition function and the process noise are both incorporated into the state transition distribution.

Although this conditional distribution is Gaussian for a fixed \(x_{n-1}\), the filtering distribution obtained by averaging over the uncertainty in \(x_{n-1}\) is generally non-Gaussian because of the nonlinear GP transition model. Consequently, the filtering distribution cannot, in general, be propagated analytically as in a linear Gaussian state-space model.

A particle filter provides a simple numerical approach to this problem \cite{DFG 2001}\cite{GSS 1993}\cite{Kitagawa 1996}. Suppose that the filtering distribution at time \(n-1\) is represented by \(M\) weighted particles,
$$
\{x_{n-1}^{(j)},w_{n-1}^{(j)}\}_{j=1}^M.
$$
In the prediction step, each particle is propagated according to the GP transition distribution:
\begin{align}
x_n^{(j)} \sim {\mathcal N}\left( \mu_f(x_{n-1}^{(j)}),s_f^2(x_{n-1}^{(j)})+q \right).
\end{align}
The propagated particles are then weighted according to the observation model. For example, when
$$
y_n=x_n+w_n, \qquad w_n\sim{\mathcal N}(0,r),
$$
the weights are updated according to
\begin{align}
w_n^{(j)} \propto p(y_n\mid x_n^{(j)}) = {\mathcal N}\bigl(y_n\mid x_n^{(j)},r\bigr).
\end{align}
The weights are then normalized, and resampling can be applied when necessary, in the same manner as in an ordinary nonlinear state-space model.

The filtering distribution is therefore approximated by the weighted empirical distribution
\begin{align}
p(x_n\mid Y_{1:n}) \approx \sum_{j=1}^{M} w_n^{(j)}\delta\bigl(x_n-x_n^{(j)}\bigr),
\end{align}
where \(\delta(\cdot)\) denotes the Dirac delta function.

Thus, once the GP transition function has been learned, its predictive distribution can be used directly as the state transition distribution in a particle filter. The GP provides a flexible data-driven representation of the unknown state dynamics, while the particle filter propagates the resulting uncertainty and estimates the latent state sequentially from noisy observations. This provides a natural connection between Gaussian process regression and nonlinear state-space modeling.

\subsection{Example: Nonlinear State-Space Model}

In this subsection, we illustrate the application of a GP-SSM using an artificially generated nonlinear time series. As the true model for generating the data, we consider the nonlinear state-space model
\begin{align}
x_n &= f(x_{n-1})+v_n, \qquad v_n\sim{\mathcal N}(0,q^2), \nonumber\\
y_n &= x_n+w_n, \hspace{15mm} w_n\sim{\mathcal N}(0,r^2),
\end{align}
where \(q^2=0.1\), \(r^2=1\), and the nonlinear transition function is
\begin{align}
f(x)=0.6x+2\sin(1.5x).
\label{Eq:true_nonlinear_transition}
\end{align}

\begin{figure}[tbp]
\begin{center}
\includegraphics[width=120mm,angle=0,clip=]{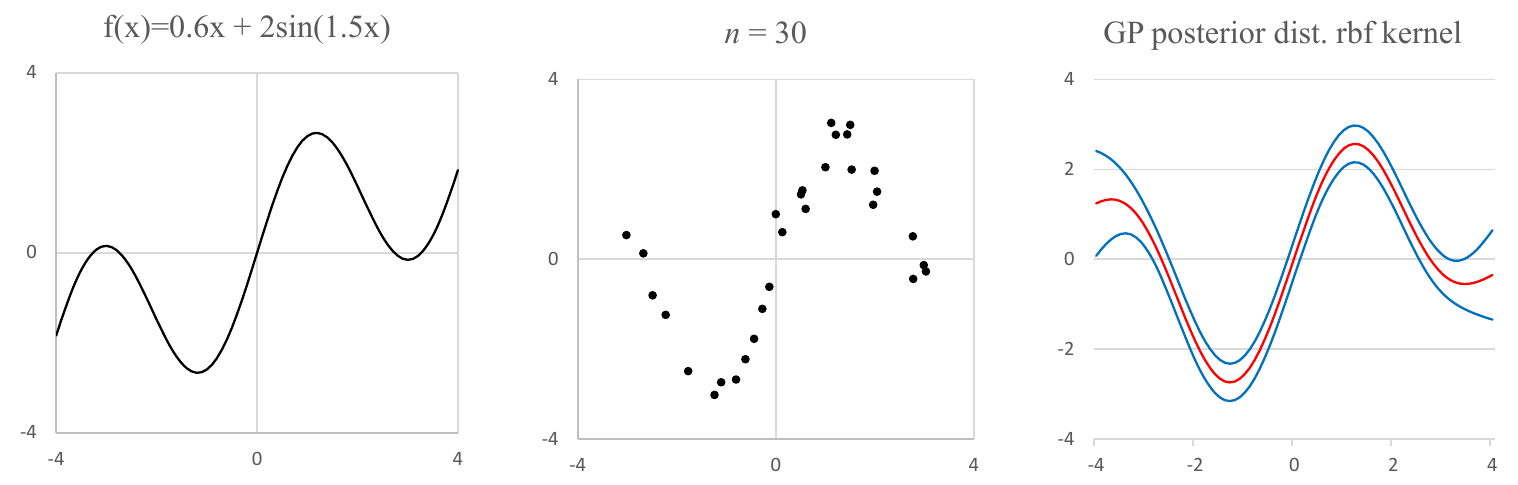}
\caption{Learning the nonlinear transition function by GP regression. The left panel shows the true transition function, the middle panel shows the \(n=30\) training observations, and the right panel shows the GP posterior distribution obtained using the rbf kernel.}
\label{Fig_GP_model_for_transition}
\end{center}
\end{figure}

The left panel of Figure \ref{Fig_GP_model_for_transition} shows the true nonlinear transition function \(f(x)\), and the middle panel shows \(n=30\) training observations generated by simulation. Table \ref{Tab:GP-regression_for_nonlinear_model} summarizes the results obtained by fitting GP regression models with the rbf kernel and the additive rbf+periodic kernel to these training data.

For the additive kernel, the period is assumed to be known and is fixed at \(2\pi/1.5\), corresponding to the period of the sine term in (\ref{Eq:true_nonlinear_transition}). The estimated parameters of the rbf component are almost identical to those obtained using the rbf kernel alone, and the two models give the same in-sample marginal log-likelihood to the precision shown in the table. Since the additive model contains two additional hyperparameters, its AIC is larger than that of the rbf model. Thus, according to AIC, the simpler rbf kernel is preferred for these training data.

Although the result is not shown in the table, we also fitted the additive model by treating the period as an unknown hyperparameter. This slightly increases the in-sample marginal log-likelihood. However, estimating the period introduces one additional hyperparameter, and the improvement in the marginal log-likelihood is not sufficient to offset the additional AIC penalty.

The right panel of Figure \ref{Fig_GP_model_for_transition} shows the posterior distribution of the transition function obtained using the rbf kernel. Near \(x=\pm4\), where few or no training observations are available, the posterior uncertainty becomes relatively large. In the region around the origin, where most of the training data are located, the posterior mean provides a good approximation to the true transition function.

\begin{table}[tbp]
\caption{Comparison of rbf and rbf$+$periodic kernels for the nonlinear transition model using \(n=30\) training observations. The table shows the in-sample marginal log-likelihood, AIC, and estimates of the kernel parameters and GP regression noise variance.}
\label{Tab:GP-regression_for_nonlinear_model}
\begin{center}
\begin{small}
\begin{tabular}{c|cc|ccccc} \hline
kernel & $L_{in}$ & AIC & $\tau_1^2$ & $\ell_1$ &$\tau_2^2$ & $\ell_2$ & $\sigma_f^2$ \rule[-5pt]{0pt}{16pt} \\ \hline
rbf           & $-27.575$ & 61.150 & 4.387 & 1.338 &   \rule[-2pt]{0pt}{14pt}&     & 0.15898 \\
rbf+periodic  & $-27.575$ & 65.150 & 4.387 & 1.338 & 0.159 & $4.872\times10^{-4}$ & 0.00006 \\
\hline
\end{tabular}
\end{small}
\end{center}
\end{table}

\begin{figure}[tbp]
\begin{center}
\includegraphics[width=120mm,angle=0,clip=]{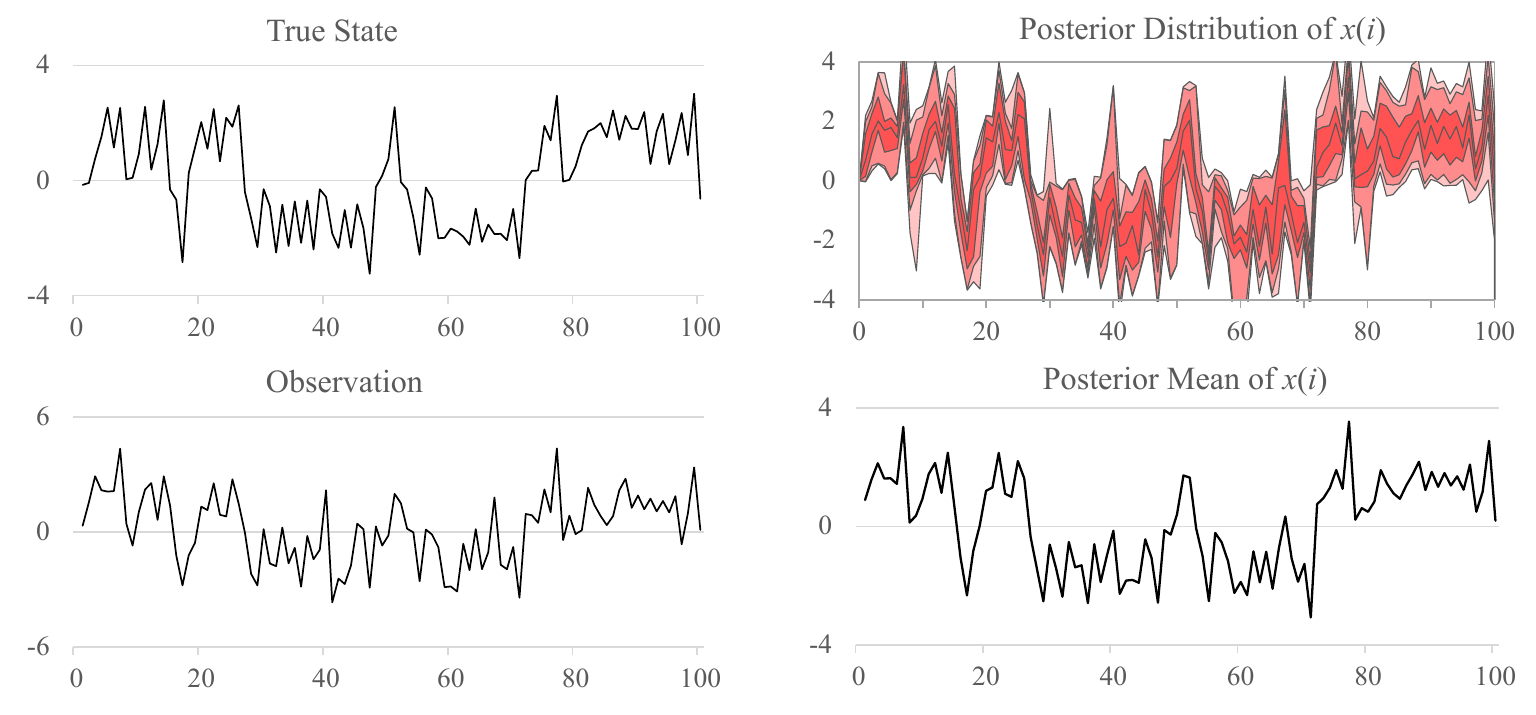}
\caption{State estimation for the nonlinear GP-SSM. The left panel shows the latent states and noisy observations generated from the true model, and the right panels show the state estimates obtained using particle filtering and fixed-lag smoothing.}
\label{Fig_GP_posterior}
\end{center}
\end{figure}

The left panel of Figure \ref{Fig_GP_posterior} shows the latent states \(x_n\) and observations \(y_n\) generated from the true nonlinear state-space model. The latent state occasionally switches between positive and negative regions, accompanied by corresponding changes in the level of the observed time series.

We next replace the known transition function in the true model by the Gaussian process learned from the training data. The resulting GP-SSM is
\begin{align}
x_n &= f(x_{n-1})+v_n, \qquad v_n\sim{\mathcal N}(0,q), \nonumber\\
y_n &= x_n+w_n, \hspace{15mm} w_n\sim{\mathcal N}(0,r),
\label{Eq:GPSSM_example}\\
f(x) &\sim {\mathcal GP}\bigl(0,k(x,x')\bigr), \nonumber
\end{align}
where \(k(x,x')\) is the fitted rbf kernel. Thus, the functional form of the nonlinear transition function in (\ref{Eq:true_nonlinear_transition}) is not used in state estimation. Instead, the transition dynamics are represented by the GP learned from the training data.

For a given particle \(x_{n-1}^{(j)}\), the GP provides the posterior distribution
\begin{align}
f(x_{n-1}^{(j)})\mid D \sim {\mathcal N}\left(\mu_f(x_{n-1}^{(j)}),s_f^2(x_{n-1}^{(j)})\right),
\end{align}
where \(D\) denotes the training data and \(s_f^2(x)\) denotes the posterior variance of the latent transition function. Taking account of the process noise, particles for the next state are therefore generated from
\begin{align}
x_n^{(j)}\sim{\mathcal N}\left(\mu_f(x_{n-1}^{(j)}),s_f^2(x_{n-1}^{(j)})+q^2 \right).
\label{Eq:GPSSM_particle_prediction}
\end{align}
The particles are then weighted according to the observation density
\begin{align}
p(y_n\mid x_n^{(j)})={\mathcal N}(y_n\mid x_n^{(j)},r^2 ).
\end{align}

The right panels of Figure \ref{Fig_GP_posterior} show the state estimates obtained from the GP-SSM using a particle filter followed by fixed-lag smoothing, with \(M=1000\) particles and a lag of 24 time steps. The upper-right panel shows the median of the fixed-lag smoothing distribution together with uncertainty bands, while the lower-right panel compares the median estimate with the true latent state. Despite the substantial observation noise and the fact that the functional form of the true transition function is not explicitly specified in the fitted model, the GP-SSM successfully tracks the major changes in the latent state, including most of the transitions between the positive and negative regions.

This simple example illustrates how a transition function learned by GP regression can be incorporated directly into a nonlinear state-space model. The GP provides a flexible, data-driven representation of the unknown dynamics, while particle filtering and smoothing propagate the resulting uncertainty and estimate the latent state from noisy observations.

\section{Concluding Remarks}

This report has presented an introductory overview of Gaussian process modeling with an emphasis on time series applications. Gaussian process regression provides a flexible, nonparametric approach in which various forms of temporal dependence can be represented through the choice of kernel functions. The numerical examples demonstrated how stationary, quasi-periodic, and seasonal time series can be modeled using individual kernels and their combinations.

The examples also illustrate the importance of evaluating GP models from both fitting and predictive viewpoints. A model that gives a larger in-sample marginal likelihood or a smaller AIC does not necessarily provide better out-of-sample prediction. In particular, the sunspot example showed that more elaborate kernels can improve the fit to the training data without improving the predictive likelihood for the evaluation data. The use of both in-sample and out-of-sample criteria can therefore provide complementary information for kernel selection and model assessment.

Composite kernels offer a convenient way of representing several types of temporal structure within a single GP model. Additive kernels are particularly useful when the underlying variation can be interpreted in terms of distinct latent components, as illustrated by the decomposition into trend, smooth local variation, and seasonality. Product kernels provide a different mechanism for constructing more complex covariance structures, such as quasi-periodic and locally periodic dependence, although their factors do not in general have the same probabilistic interpretation as the components of an additive kernel.

Finally, the GP-SSM example demonstrated that Gaussian processes can also be used within the state-space modeling framework to represent unknown nonlinear transition functions. Once a transition function has been learned from data, its predictive distribution can be incorporated into particle filtering and smoothing procedures for latent-state estimation. This provides a natural connection between Gaussian process regression and nonlinear state-space modeling, and extends the applicability of GP methods beyond direct modeling of observed time series.

\end{document}